\documentclass[twocolumn]{revtex4}
\pdfoutput=1
\usepackage{graphicx}
\usepackage{dcolumn}
\usepackage{longtable}
\usepackage{amsmath}
\usepackage{amssymb}

\newcommand{\tbeta}{\tilde{\beta}}
\newcommand{\tK}{\tilde K}

\begin{document}

\title
{Bohr Hamiltonian with deformation-dependent mass term
for the Kratzer potential}

\author
{Dennis Bonatsos$^1$, P. E. Georgoudis$^1$, N. Minkov$^2$, D. Petrellis$^1$, and C. Quesne$^3$}

\affiliation
{$^1$Institute of Nuclear and Particle Physics, National Centre for Scientific Research 
``Demokritos'', GR-15310 Aghia Paraskevi, Attiki, Greece}

\affiliation
{$^2$Institute of Nuclear Research and Nuclear Energy, Bulgarian Academy of Sciences, 72 Tzarigrad Road, 1784 Sofia, Bulgaria}

\affiliation
{$^3$ Physique Nucl\'eaire Th\'eorique et Physique Math\'ematique, Universit\'e Libre de Bruxelles, Campus de la Plaine CP229,
Boulevard du Triomphe, B-1050 Brussels, Belgium}

\begin{abstract}

The Deformation Dependent Mass (DDM) Kratzer model is constructed by considering the Kratzer potential in a Bohr Hamiltonian, 
in which the mass is allowed to depend on the nuclear deformation, and solving it by using techniques of supersymmetric quantum mechanics (SUSYQM),
involving a deformed shape invariance condition. Analytical expressions for spectra and wave functions are derived for separable potentials in the cases of $\gamma$-unstable nuclei, axially symmetric prolate deformed nuclei, and triaxial nuclei, implementing the usual approximations in each case. 
Spectra and $B(E2)$ transition rates are compared to experimental data. The dependence of the mass on the deformation, dictated by SUSYQM for the potential used, moderates the increase of the moment of inertia with deformation, removing a main drawback of the model. 

\end{abstract}


\maketitle
    
\section{Introduction}

In the non-relativistic Schr\"odinger equation usually the 
mass is assumed to be independent of position. However, there
are certain physical systems in which, according to the experimental
data, the effective mass should be position dependent. Apart from semiconductor
theory, in which effective masses depending on the spatial coordinates have been used 
since many years \cite{BenDaniel,Gora,Bastard,Zhu,Roos,Morrow}, 
this is also the case in atomic nuclei, especially in the collective model of Bohr \cite{Bohr}.

If the mass is position dependent, then it does not commute with the momentum. 
Therefore there are many ways to generalize the usual form of the kinetic energy 
in order to obtain a Hermitian operator. In Ref. \cite{Roos}, the most general non-relativistic 
Schr\"odinger equation which possesses a position dependent mass and at the same time 
respects hermiticity was proposed. A further step was taken in Refs. \cite{QT4267,Q2929},
in which a general scheme was proposed, through the introduction of a parameter $a$ 
in the kinetic energy term, which apart from hermiticity also ensures the 
exact solvability of the relevant non-relativistic Schr\"odinger equation, 
in which the mass dependence on the position is reflected in the parameter $a$. 

In the realm of nuclear physics, the need for three different mass coefficients 
in the low lying collective bands of well-deformed axially symmetric even-even nuclei 
has been demonstrated using the experimental data for spectra and $B(E2)$ transition rates 
by Jolos and von Brentano \cite{Jolos76,Jolos77,Jolos78}. In an extended version 
of this approach \cite{Jolos79,Jolos80} a mass tensor with deformation dependent 
components is introduced, the method being applicable to nuclei of arbitrary shape. 
This approach has been recently extended to odd nuclei \cite{Erma84,Erma85}. 

It has been recently proved \cite{DDMD} that by allowing the nuclear mass 
to depend on the deformation, the rate of increase of the moment of inertia with deformation 
is moderated, thus removing a main drawback \cite{Ring} of the Bohr Hamiltonian \cite{Bohr}. 
This has been achieved by using a Davidson potential \cite{Dav} and taking advantage 
of supersymmetric quantum mechanics (SUSYQM) techniques \cite{SUSYQM1,SUSYQM2} 
developed in the study of quantum systems with mass depending on the coordinates 
\cite{QT4267,Q2929,Q13107}.

It should be noticed that the Davidson potential \cite{Dav}, initially introduced 
for the description of molecular spectra, has been used for the description of nuclei 
since long ago \cite{Rohoz,EEP}. A major step forward has been the clarification 
of its group theoretical structure when used within the Bohr Hamiltonian, 
which was proved to be SU(1,1)$\times$SO(5) \cite{Bahri}. This breakthrough 
led to the development of the algebraic collective model, 
a very rapidly converging approach allowing the calculation of spectra 
and transition probabilities of nuclei of any shape \cite{Rowe735,Rowe753,Caprio,Welsh}. 

In the present work we consider the deformation dependent mass (DDM) Bohr Hamiltonian with a Kratzer potential \cite{Kratzer}.
This potential, also used initially in molecular physics, has been first used for the description of nuclei by Fortunato and Vitturi \cite{FV1,FV2}.
We are motivated by the following questions.

(i) It is known that in the SUSYQM framework \cite{SUSYQM1,SUSYQM2,Q2929} the functional form of the dependence of the mass 
on the deformation is different for each potential. A first challenge is to prove that a DDM solution exists for the Bohr Hamiltonian 
with a Kratzer potential, and to find the form of this functional dependence.   

(ii) The shapes of the Davidson \cite{Dav} and Kratzer \cite{Kratzer} potentials, shown in Fig.~1, are similar for small values of $\beta$,
where the $1/\beta^2$ term dominates in both cases, but they differ substantially at large values of $\beta$, where the tails of the wave functions 
behave as $e^{-\beta^2/2}$ \cite{ESDPRC} and $e^{-\beta/2}$ \cite{FV1,FV2} respectively.
It is interesting to examine to what extent numerical results for spectra and $B(E2)$ transition rates are 
influenced by the change in the potential. The level spacings within the $\beta_1$ band, which are usually overestimated
by collective models based on the Bohr Hamiltonian, including the DDM Davidson model, are of particular interest.  

\begin{figure*}[hbt]
\includegraphics[width=1\textwidth]{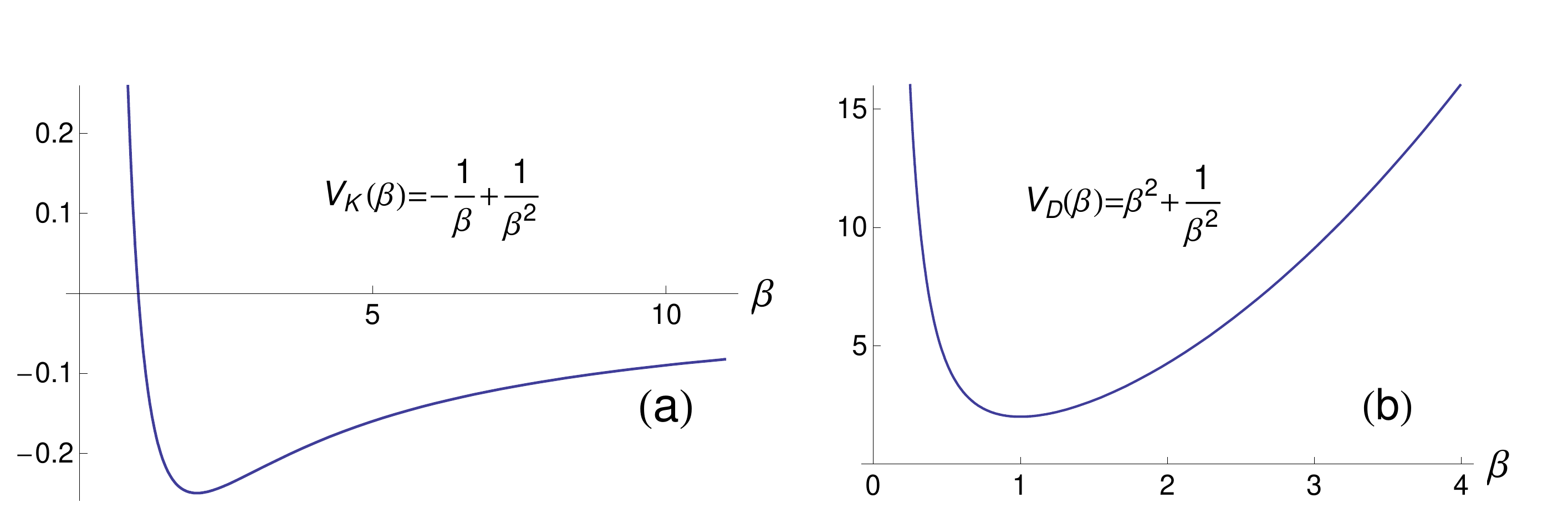}
\caption{(Color online) The Kratzer (left) and Davidson (right) potentials. The quantities shown are dimensionless, while 
all free parameters have been set equal to unity, for the sake of simplicity. }
\end{figure*}

(iii) The DDM Bohr Hamiltonian with the Davidson potential works well for deformed nuclei, but fails in describing 
the nuclei lying at the critical point between the spherical and deformed regions \cite{McCutchan,RMP82}, known to be examples of
the X(5) critical point symmetry \cite{IacX5}. It is interesting to examine if the DDM Bohr Hamiltonian with the Kratzer potential
overcomes this drawback.   

In Section II the DDM Bohr Hamiltonian is briefly reviewed, while in Section III the special case of the Kratzer potential is considered.
Analytical expressions for spectra and wave functions are given in Sections IV and V respectively, while the calculation of $B(E2)$ 
transition rates is described in Section VI. Numerical results for spectra and $B(E2)$ transition rates are given in Section VII, 
while Section VIII contains the conclusions and plans for further work. The use of the deformed shape invariance principle for the construction 
of the spectrum is given in Appendix 1, while many technical details concerning the wave functions are given in Appendices 2-6.  Finally, in Appendix 7,
scaling factors are discussed. 

\section{Bohr Hamiltonian with deformation-dependent mass} 

The original Bohr Hamiltonian \cite{Bohr} is
\begin{eqnarray}\label{eq:e1}
H_B = -{\hbar^2 \over 2B} \left[ {1\over \beta^4} {\partial \over \partial 
\beta} \beta^4 {\partial \over \partial \beta} + {1\over \beta^2 \sin 
3\gamma} {\partial \over \partial \gamma} \sin 3 \gamma {\partial \over 
\partial \gamma} \right. \nonumber \\
\left. - {1\over 4 \beta^2} \sum_{k=1,2,3} {Q_k^2 \over \sin^2 
\left(\gamma - {2\over 3} \pi k\right) } \right] +V(\beta,\gamma),
\end{eqnarray}
where $\beta$ and $\gamma$ are the usual collective coordinates, while
$Q_k$ ($k=1$, 2, 3) are the components of angular momentum in the intrinsic 
frame, and $B$ is the mass parameter, which is usually considered constant.  

Allowing the mass to depend on
the deformation coordinate $\beta$ (which measures departure from spherical shape),
\begin{equation}\label{massdep}
B(\beta)=\frac{B_0}{(f(\beta))^2}, 
\end{equation}
where $B_0$ is a constant and $f$ a function of $\beta$ only, the Bohr equation becomes \cite{DDMD}
\begin{equation}
\begin{split}
\label{eq:mBohr}
 & H \Psi = \left[ 
-{1\over 2} {\sqrt{f}\over \beta^4} {\partial \over \partial \beta} 
\beta^4 f {\partial \over \partial \beta} \sqrt{f}
-{f^2 \over 2 \beta^2 \sin 3\gamma} {\partial \over \partial \gamma} 
\sin 3\gamma {\partial \over \partial \gamma} \right.  \\
 & \left. + {f^2\over 8 \beta^2} 
\sum_{k=1,2,3} {Q_k^2 \over \sin^2\left(\gamma -{2\over 3} \pi k \right)}
+ v_{eff} \right] \Psi = \epsilon \Psi,
\end{split}  
\end{equation}
where reduced energies 
\begin{equation}\label{reducedE}
\epsilon = B_0 E/\hbar^2
\end{equation}
and reduced potentials
\begin{equation}\label{reducedV}
v= B_0 V/\hbar^2
\end{equation} 
have been used, and \cite{DDMD}
\begin{eqnarray}
v_{eff}= v(\beta,\gamma)+ {1\over 4 } (1-\delta-\lambda) f \nabla^2 f  \nonumber \\
+ {1\over 2} \left({1\over 2} -\delta\right) \left( {1\over 2} -\lambda\right)
(\nabla f)^2 ,
\end{eqnarray}
where $\delta$ and $\lambda$ are free parameters, stemming from the following cause. 
If the mass is position dependent, then it does not commute with the momentum. 
Therefore there are many ways to generalize the usual form of the kinetic energy 
in order to obtain a Hermitian operator. It was actually in Ref. \cite{Roos} where it was proved 
that the most general form of such a Hermitian Hamiltonian contains two free parameters 
(denoted by $\delta$ and $\lambda$ in the present work). In Section VII it will be seen 
that these parameters play practically no role in the present work.  

Exact separation of variables can be achieved in the usual three cases.

a) $\gamma$-unstable nuclei, in which the potential $v(\beta,\gamma)$ depends only 
on the variable $\beta$, i.e. $v(\beta)=u(\beta)$ \cite{Wilets,IacE5}, 
and the wave functions are of the form 
\begin{equation}
\Psi (\beta, \gamma, \theta_i)= \xi(\beta) \Phi(\gamma, \theta_i),
\end{equation}
where $\theta_i$ ($i=1$, 2, 3) are the Euler angles.
  
b) Axially symmetric prolate deformed nuclei with $\gamma \approx 0$, in which the potential 
is assumed to be of the form \cite{Wilets,F2,F1,F3,ESDPRC}
\begin{equation}\label{eq:e7b}
v(\beta,\gamma)= u(\beta)+{f^2 \over \beta^2} w(\gamma),
\end{equation}
with 
\begin{equation}\label{wgamma}
w(\gamma)= {1\over 2} (3c)^2 \gamma^2 ,
\end{equation}
where $c$ is a free parameter,
while the wave functions read \cite{IacX5} 
\begin{equation}
\Psi(\beta,\gamma,\theta_i)=   \xi_L(\beta) \eta_K(\gamma)
{\cal D}^L_{M,K}(\theta_i),
\end{equation}
where ${\cal D}(\theta_i)$ denote Wigner functions of the Euler angles, 
$L$ is the angular momentum quantum number, while $M$ and $K$ are the 
quantum numbers of the projections of angular momentum on the laboratory-fixed 
$z$-axis and the body-fixed $z'$-axis respectively. 

c) Triaxial nuclei with $\gamma \approx \pi/6$, in which again the potential is assumed to be 
of the form of Eq. (\ref{eq:e7b}), 
but  with 
\begin{equation}\label{wgammatriax}
w(\gamma)= {1\over 4} c \left( \gamma -{\pi\over 6}\right)^2  ,
\end{equation}
while the wave functions read \cite{Z5}
\begin{equation}
\Psi(\beta,\gamma,\theta_i)= \xi_{L,\alpha}(\beta) \eta(\gamma)
{\cal D}^L_{M,\alpha}(\theta_i),
\end{equation}
where ${\cal D}(\theta_i)$ denote Wigner functions of the Euler angles, 
$L$ is the angular momentum quantum number, while $M$ and $\alpha$ are the 
quantum numbers of the projections of angular momentum on the laboratory-fixed 
$z$-axis and the body-fixed $x'$-axis respectively. 

The angular equations and wave functions can be found in Ref. \cite{DDMD}. The radial 
equations in all three cases take the common form \cite{DDMD}
\begin{equation}\label{eq:e14}
H R(\beta)= - \left(\sqrt{f}{d\over d\beta}\sqrt{f}\right)^2 R(\beta) + 2 u_{eff} R(\beta) 
=2 \epsilon R(\beta), 
\end{equation}
where 
\begin{equation}\label{xiR}
\xi(\beta)= {R(\beta)\over \beta^2}, 
\end{equation}
\begin{eqnarray}\label{eq:e15}
u_{eff}= u + {1\over 4} (1-\delta-\lambda)  f \left( {4 f'\over \beta} 
+f''\right) \nonumber \\ 
+ {1\over 2} \left( {1\over 2}-\delta\right) 
\left( {1\over 2}-\lambda\right) (f')^2 + {f^2+\beta f f'\over \beta^2}+ {f^2 \over 2 \beta^2}  \Lambda,    
\end{eqnarray}
$f'$ ($f''$) denote the first (second) derivative of $f$, 
while $\Lambda$ in each case acquires the following form.

a) For $\gamma$-unstable nuclei 
\begin{equation}\label{Ltau}
 \Lambda=\tau(\tau+3),
\end{equation}
with $\tau$ being the seniority quantum number \cite{Bes}. The values of angular momentum $L$ occurring 
for each $\tau$ are provided by a well known algorithm and are listed in \cite{IA,Wilets}.
Within the ground state band (gsb) one has $L=2\tau$.
  
b) For axially symmetric prolate deformed nuclei,
\begin{equation}\label{Ltilde}
 \Lambda = {L(L+1)-K^2\over 3}+ (6c) (n_\gamma+1),
\end{equation}
where $n_\gamma$ is the quantum number related to $\gamma$-oscillations. 

c) For triaxial nuclei with $\gamma \approx \pi/6$,
\begin{equation}\label{Lbar2}
 \Lambda = {L(L+4)+3 n_w(2L-n_w) \over 4} + \sqrt{2c} \left(n_\gamma+{1\over 2}\right),
\end{equation}
where $n_w=L-\alpha$ is the wobbling quantum number \cite{BM,MtV}. 

\section{The Kratzer potential} 

Up to now no assumption about the specific form of the potential 
$u(\beta)$ and the deformation function $f(\beta)$ has been made.
From the results for 3-dimensional systems reported 
in Ref. \cite{Q2929}, we know that for each potential a different 
deformation function is appropriate. 

In Ref. \cite{DDMD}, the 
Davidson potential \cite{Dav}
\begin{equation} \label{eq:e16}
u(\beta)=\beta^2 + {\beta_0^4\over \beta^2},
\end{equation}
where the parameter $\beta_0$ indicates the position of the minimum 
of the potential, has been considered in  the framework of the Bohr Hamiltonian,
the appropriate deformation function being 
\begin{equation}\label{eq:e17}
f(\beta)=1+a \beta^2, \qquad a \ll 1.
\end{equation}

Here we are going to consider the Kratzer potential \cite{Kratzer} 
\begin{equation}\label{Kratz}
  u(\beta) = - \frac{1}{\beta} + \frac{\tilde{B}}{\beta^2},
\end{equation}
for which the deformation function is expected \cite{Q2929} to be 
\begin{equation}\label{deffun}
  f(\beta) = 1 + a \beta, \qquad a \ll 1. 
\end{equation}

Using these forms for the potential and the deformation function 
in Eq. (\ref{eq:e15}) one obtains 
\begin{multline}
   2 u_{\rm eff} = k_0 + \frac{k_{-1}}{\beta} + \frac{k_{-2}}{\beta^2}, \\
   k_0 = a^2 [2(3 - \delta - \lambda) + \tfrac{1}{4} (1 - 2\delta) (1 - 2\lambda) + \Lambda], \\
   k_{-1} = - 2 + 2a [(4 - \delta - \lambda) + \Lambda], \\
   k_{-2} = 2 + \Lambda + 2 \tilde{B}, 
 \label{eq:ueff}
\end{multline}
where $a$ is the deforming parameter, $\delta$ and $\lambda$ are free parameters coming from the 
construction procedure of the kinetic energy term \cite{Roos} and going to be discussed further in Section VII.A, 
and $\Lambda$ is the eigenvalue coming from the 
exact separation of variables, given by Eqs. (\ref{Ltau}), (\ref{Ltilde}), (\ref{Lbar2}), depending on the nature of the nucleus in discussion
($\gamma$-unstable, axially symmetric prolate deformed, triaxial with $\gamma \approx \pi/6$ respectively). 

The radial equation for the Kratzer potential is solved in Appendix 1, 
using deformed shape invariance. 

\section{Energy spectrum}\label{enespe}

Using the results of Appendix 1, 
the energy spectrum of Eq.~(\ref{eq:e14}) is given by
\begin{multline}
  \epsilon_n = \frac{1}{2} \left[k_0 - \left(\frac{k_{-1} + a \left[n^2 + \frac{1}{2}(1 + \Delta)(2n+1)\right]}
       {2n+1+\Delta}\right)^2\right], \\
        \qquad n=0, 1, 2, \ldots, \label{eq:energy}
\end{multline}
where
\begin{equation}
\Delta \equiv \sqrt{1 + 4 k_{-2}}.
\end{equation}

Equation (\ref{eq:energy}) only provides a formal solution to the bound-state energy spectrum. The range of $n$ values is actually determined by the existence of corresponding physically acceptable wave 
functions, the relevant conditions being stated in Appendix 2.

In Eq. (\ref{eq:energy}) the quantities $k_0$, $k_{-1}$, $k_{-2}$ are given by Eq. (\ref{eq:ueff}), 
in which $\Lambda$ is given by Eq.(\ref{Ltau}), (\ref{Ltilde}), or (\ref{Lbar2}), 
for $\gamma$-unstable, axially symmetric prolate deformed, or triaxial nuclei respectively. 
The ground state band is obtained for $n=0$, while for $n=1$ and $n=2$ the quasi-$\beta_1$ and quasi-$\beta_2$ bands 
are obtained respectively. 

In the limit of no dependence of the mass on the deformation, i.e., $a\to0$, one has from Eq. (\ref{eq:ueff})
\begin{equation}
k_0=0, \qquad k_{-1}=-2.
\end{equation}

In this limit the spectrum becomes 
\begin{equation}
\epsilon_n =-{1\over 2 \left(n+{1\over 2} + {\Delta\over 2}  \right)^2}.  
\end{equation}

For $\gamma$-unstable nuclei this reads 
\begin{equation}
\epsilon_n =-{1\over 2 \left(n+{1\over 2} +  \sqrt{ \left(\tau+{3\over 2} \right)^2+2\tilde B} \right)^2},  
\end{equation}
in agreement with Eq. (11) of Ref. \cite{FV1}.

In the axially symmetric prolate deformed case the energies for the ground state and $\beta$ bands 
in the limit $a\to 0$ read
\begin{equation}
\epsilon_n =-{1\over 2 \left(n+{1\over 2} +\sqrt{{9\over 4}+ {L(L+1)\over 3}+2 \tilde B }  \right)^2},  
\end{equation}
in agreement with Eq. (10) of Ref. \cite{FV2}.

\section{Wave functions}

The ground-state wavefunction is found in Appendix 3 to read 
\begin{equation}\label{RR0}
  R_0(\beta) = N_0 \beta^{-\mu} 
  f^{\frac{1}{2} \left(\mu + \frac{\tK}{\mu} - 1\right)},
\end{equation}
where 
\begin{equation}
\mu= -{1\over 2}(1+\Delta), 
\end{equation}
while the normalization factor $N_0$ is given by Eq. (\ref{Nzero}), 
and $\tilde K$ is given by Eq. (\ref{eq:K}).  

As discussed in Appendix 3, this wave function is physically acceptable if  the inequality 
\begin{equation}
  |\mu| = \frac{1}{2} (1+\Delta) < - \frac{k_{-1}}{a}, 
\end{equation}
 is satisfied by the parameters of the problem.
 
The excited state wave functions are found in Appendix 4 to be 
\begin{multline}\label{RRn}
  R_n(\beta) = N_n \beta^{- \mu_n} f^{\frac{1}{2} \left(\mu_n + \frac{\tK}{\mu_n} - 1\right)} 
  P_n^{\left(\mu_n - \frac{\tK}{\mu_n}, \mu_n + \frac{\tK}{\mu_n}\right)}(t), \\
  \qquad t = 
  \frac{2+a\beta}{a\beta},
\end{multline}
where $\mu_n=\mu-n$, by $P_n^{(\alpha,\beta)}(t)$ the Jacobi polynomials \cite{AbrSte} are denoted,  
while the normalization coefficient $N_n$ is given by Eq. (\ref{Nall}) in Appendix 5. 

The reduction of the present wave functions to the form they have in the $a\to 0$ limit \cite{FV1,FV2}
is carried out in Appendix 6. 

\section{$B(E2)$ transition rates}

For the calculation of $B(E2)$ transition rates, the formulae given in Section X of Ref. \cite{DDMD}
apply, with the wave functions $R_n(\beta)$ being replaced by the present results.  

\section{Numerical results} 

\subsection{Spectra of $\gamma$-unstable nuclei}\label{specgam}

Rms fits of spectra have been performed, using the quality measure 
\begin{equation}\label{eq:e99}
\sigma = \sqrt{ { \sum_{i=1}^n (E_i(exp)-E_i(th))^2 \over
(n-1)E(2_1^+)^2 } }.
\end{equation}
The same set of experimental data for spectra and $B(E2)$ transition rates has been used 
as in the cases of the Deformation Dependent Mass (DDM) Davidson model \cite{DDMD},
the Exactly Separable Davidson (ESD) model \cite{ESDPRC}, and the Morse potential 
\cite{MorseI,MorseII}, in order to facilitate comparisons of the various models among themselves 
and to the data. 

The theoretical predictions for the levels are obtained from Eq. (\ref{eq:energy}), the ground state band corresponding 
to $n=0$ and the quasi-$\beta_1$ band having $n=1$.
The levels of the quasi-$\gamma_1$ band are obtained through their degeneracies to members of the ground state band,
implied by the SO(5)$\supset$SO(3) reduction rules \cite{Wilets,IA}, also listed in Table I of Ref. \cite{BonE5}. 
Within the ground state band (gsb) one has $L=2\tau$. The $L=2$ member of the quasi-$\gamma_1$ band 
is degenerate with the $L=4$ member of the gsb, the $L=3$, 4 members of the quasi-$\gamma_1$ band 
are degenerate to the $L=6$ member of the gsb, the $L=5$, 6 members of the quasi-$\gamma_1$ band 
are degenerate to the $L=8$ member of the gsb, and so on.  
 
The results shown in Table I have been obtained for $\delta=\lambda=0$. 
One can easily verify that 
different choices for $\delta$ and $\lambda$ lead to a renormalization of the parameter values $a$ and $\beta_0$, 
the predicted energy levels remaining exactly the same. 

Concerning the physical content of the parameter $a$, it is instructive to consider in detail in Table~I the Xe isotopes,
known \cite{Casten} to lie in the $\gamma$-unstable region and already discussed in the framework of the DDM Davidson model \cite{DDMD} . 
They  extend from the borders of the neutron shell (with $^{134}$Xe$_{80}$ lying just below the N=82 shell closure)
to the midshell ($^{120}$Xe$_{66}$) and even beyond, exhibiting increasing collectivity (increasing $R_{4/2}=E(4_1^+)/E(2_1^+)$ ratios)
from the border to the midshell.

Moving from the border of the neutron shell to the midshell, the following remarks apply.

i) The Kratzer parameter $\tilde B$ is increasing smoothly as one moves from the border towards the middle of the shell.

ii) The $a$ parameter, expressing the dependence of the mass on the deformation, is zero, or close to zero, 
for the first 4 isotopes close to the border, while it acquires substantially non-zero values for the last 5 isotopes 
close to mid-shell. This indicates that for nearly spherical nuclei no dependence of the mass on the deformation 
is needed, while it is becoming necessary as soon as substantial deviations from the spherical shape set in. 

\begin{figure*}[hbt]
\includegraphics[width=1\textwidth]{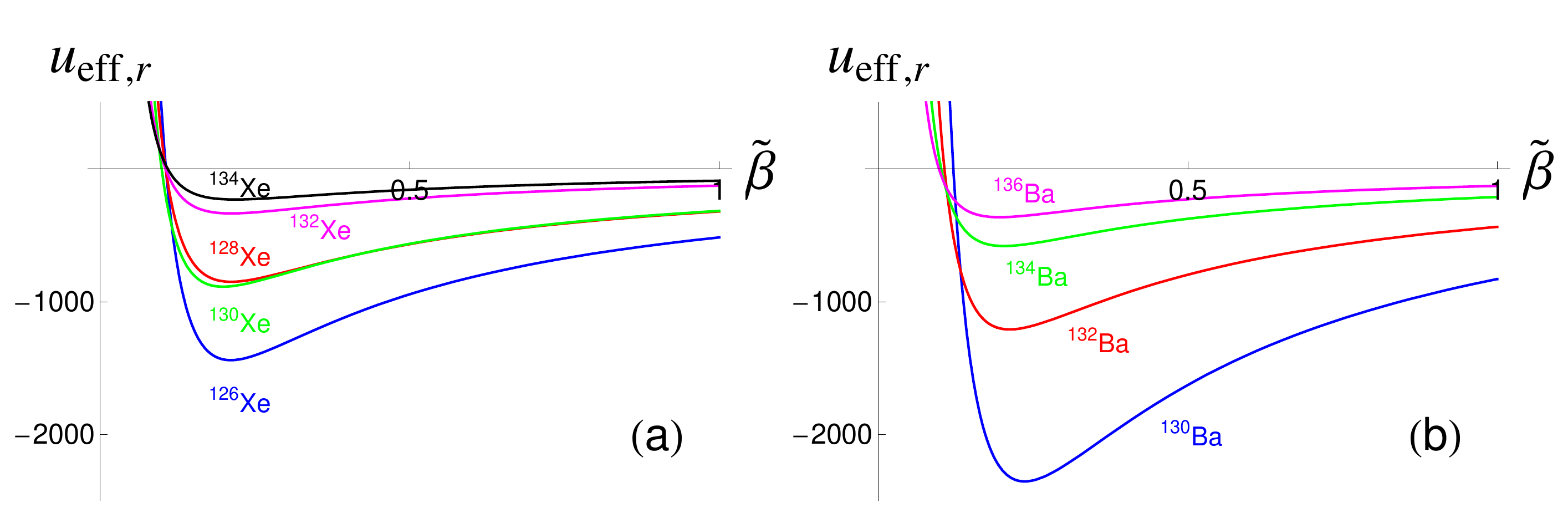}
\caption{(Color online) Effective potentials (Eq. (\ref{eq:ueff})) for $L=4$ for some Xe (a) and Ba (b) isotopes, corresponding to the 
parameters of Table I. The quantities shown are dimensionless. Rescaling has been carried out, according to Appendix 7, using the rescaling parameters $A$ given in Table VII. 
The rescaled effective potentials are given by Eq. (\ref{eff3}), while the rescaled abscissa, $\tilde \beta$, is given by Eq. (\ref{tildeb}).  }
\end{figure*}

The effective potentials of Eq. (\ref{eq:ueff}) for some Xe isotopes, appropriately rescaled as described in Appendix 7, 
are shown in Fig. 2(a).  It is clear that the potentials 
get deeper as one moves from the border of the neutron shell ($^{134}$Xe$_{80}$) to the midshell ($^{120}$Xe$_{66}$).
 
Other chains of isotopes also show similar behavior. As an example, the effective potentials of Eq. (\ref{eq:ueff}) for some Ba isotopes,
appropriately rescaled as described in Appendix 7, are shown in Fig. 2(b).
It is again clear that the potentials get deeper as one moves from the border of the neutron shell ($^{136}$Ba$_{80}$) towards the midshell ($^{122}$Ba$_{66}$).

\begin{figure*}[hbt]
\includegraphics[width=1\textwidth]{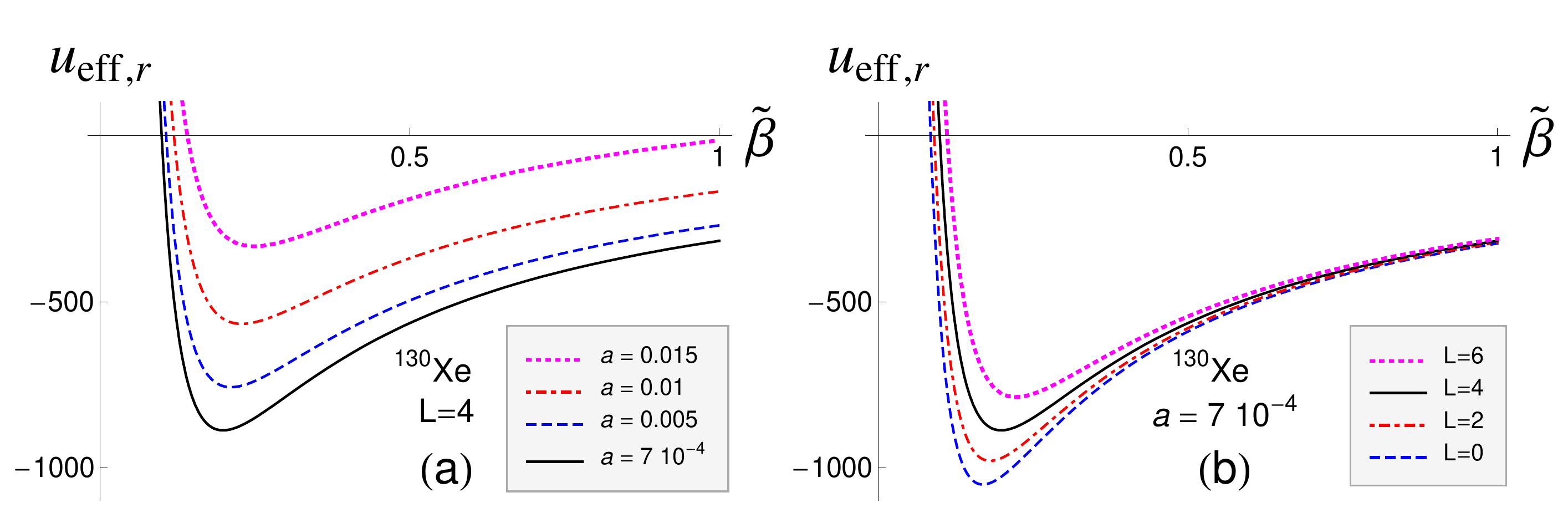}
\caption{(Color online) Dependence of the effective potentials of Eq. (\ref{eq:ueff}) on the parameter $a$ [panel (a)] 
and on the angular momentum $L$ [panel (b)]. The quantities shown are dimensionless. The effective potential for the $L=4$ state of the ground state band of $^{130}$Xe 
(corresponding to the parameters of Table I), plotted in both panels, is used as the basis of the comparison.
Rescaling has been carried out, according to Appendix 7, using the rescaling parameters $A$ given in Table VII. 
The rescaled effective potentials are given by Eq. (\ref{eff3}), while the rescaled abscissa, $\tilde \beta$, is given by Eq. (\ref{tildeb}).   }
\end{figure*}

The dependence of the effective potentials of Eq. (\ref{eq:ueff}) on the parameter $a$ and on the angular momentum $L$ 
is shown in Fig. 3, after appropriate rescaling according to Appendix 7. The effective potential of the $L=4$ state of the ground state band of $^{130}$Xe is used as the basis
for the comparison. It is seen that the effective potential becomes less deep as the parameter $a$ is increased. 
It also becomes less deep as the angular momentum $L$ is increased. 

It is remarkable that all effective potentials shown in Figs. 2 and 3 look qualitatively similar to the pure Kratzer potential of Fig. 1(left).

\subsection{Spectra of axially symmetric deformed nuclei}\label{specdef}

Fits of spectra of deformed rare earth and actinide nuclei are shown in Table~II. 
The energy levels are obtained from Eq. (\ref{eq:energy}). The ground state band is obtained for $n=0$ and 
the $\beta_1$ band for $n=1$, while both have $n_\gamma=0$ and $K=0$. 
The $\gamma_1$ band is obtained for $n=0$, $n_\gamma=1$ and $K=2$. 
Again, the choice $\delta=\lambda=0$ has been made, and it is seen that 
different choices for $\delta$ and $\lambda$ lead to a renormalization of the parameter values $a$, $B$, and $c$,  
the predicted energy levels remaining exactly the same. 

The quality of the fits obtained can also be seen in Table~III, where the calculated energy levels 
of $^{170}$Er and $^{232}$Th are compared to experiment. 

The main discrepancy between theory and experiment in the case of the DDM Davidson \cite{DDMD} 
was found in  the $\beta_1$-bands, in which the theoretical level spacings were larger than the experimental ones. 
This was attributed to the shape of the Davidson potential, which raises to infinity at large $\beta$, 
pushing $\beta$ bands higher and increasing their interlevel spacing. It is known that this problem can be avoided
by using a potential going to some finite value at large $\beta$ \cite{finitew}, like the Morse potential \cite{Morse}. 
The Kratzer potential is going to zero for large $\beta$, thus avoiding the problem 
of the overestimation of the level spacings within the $\beta_1$ band, as seen clearly in Table III. 

\begin{figure*}[hbt]
\includegraphics[width=1\textwidth]{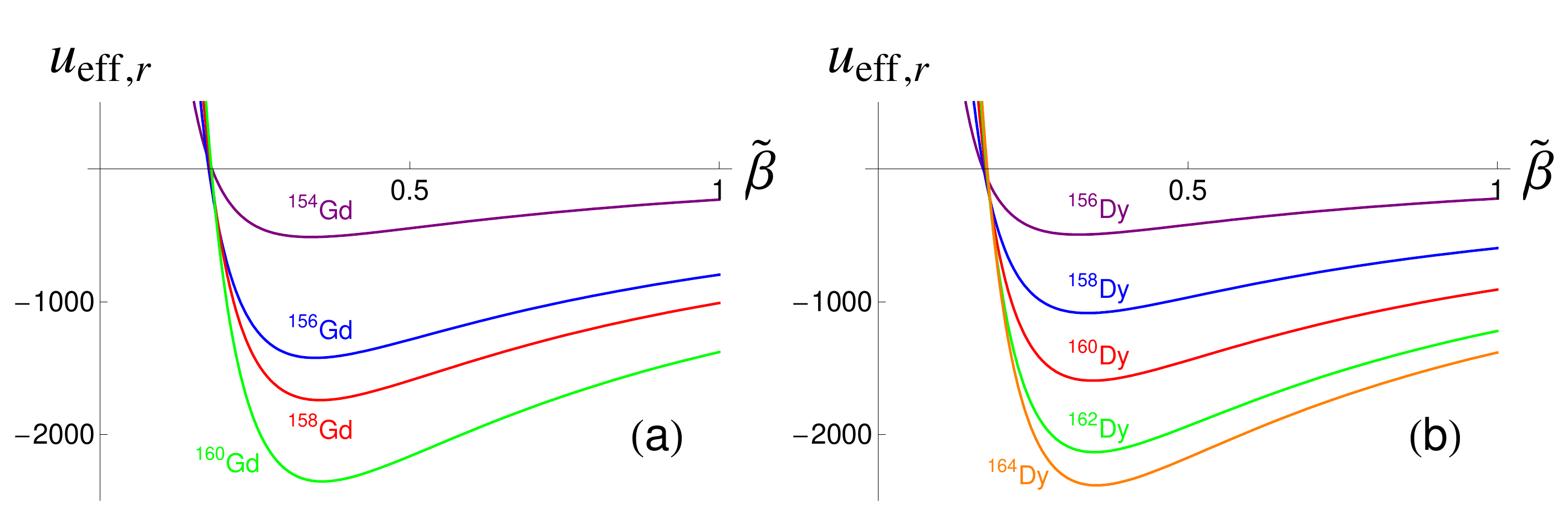}
\caption{(Color online) Effective potentials (Eq. (\ref{eq:ueff})) for $L=4$ for some Gd (a) and Dy (b) isotopes, corresponding to the 
parameters of Table II.  The quantities shown are dimensionless. Rescaling has been carried out, according to Appendix 7, using the rescaling parameters $A$ given in Table VIII. 
The rescaled effective potentials are given by Eq. (\ref{eff3}), while the rescaled abscissa, $\tilde \beta$, is given by Eq. (\ref{tildeb}).
 }
\end{figure*}

In Fig. 4 the effective potentials of Eq. (\ref{eq:ueff}) for some Gd and Dy isotopes,
appropriately rescaled as described in Appendix 7,  are shown. It is clear that the potentials 
get deeper as one moves from the border of the neutron shell towards the midshell.

Another difference between the DDM Davidson and the present DDM Kratzer model is that the former 
cannot describe the ${\rm N}=90$ isotones $^{150}$Nd, $^{152}$Sm, $^{154}$Gd, and $^{156}$Dy, which are 
considered \cite{RMP82} as the best examples of the X(5) \cite{IacX5} critical point symmetry, while in the latter
a good description is obtained. Indeed, in Table II of Ref. \cite{DDMD} one sees that large $\sigma$ deviations 
are obtained in the DDM Davidson case, while the spectra obtained in the present DDM Kratzer case are reported 
in Table IV, along with the parameter-free X(5) predictions \cite{IacX5,BonX5,Bijker}. In the $\beta_1$ bands, in particular, 
we see that the present approach, using the same number of parameters (three in the case of axially deformed nuclei) 
as the DDM Davidson model, avoids the overestimation of the interlevel spacings. 

The ability of the DDM Kratzer model to describe the $N=90$ isotones, in which the DDM Davidson model fails,
can be understood by considering the shapes of the two potentials. The Kratzer potential for appropriate parameter values 
can acquire the shape of a deep well, thus resembling the infinite square well potential used in the X(5) model \cite{IacX5}, 
known to describe these ${\rm N}=90$ isotones. 

In Fig. 5, the (rescaled according to Appendix 7) effective potentials for the ${\rm N}=90$ nuclei 
being good examples of the X(5) critical point symmetry are isolated, corroborating the remarks made above.  

\begin{figure}[hbt]
\includegraphics[width=0.45\textwidth]{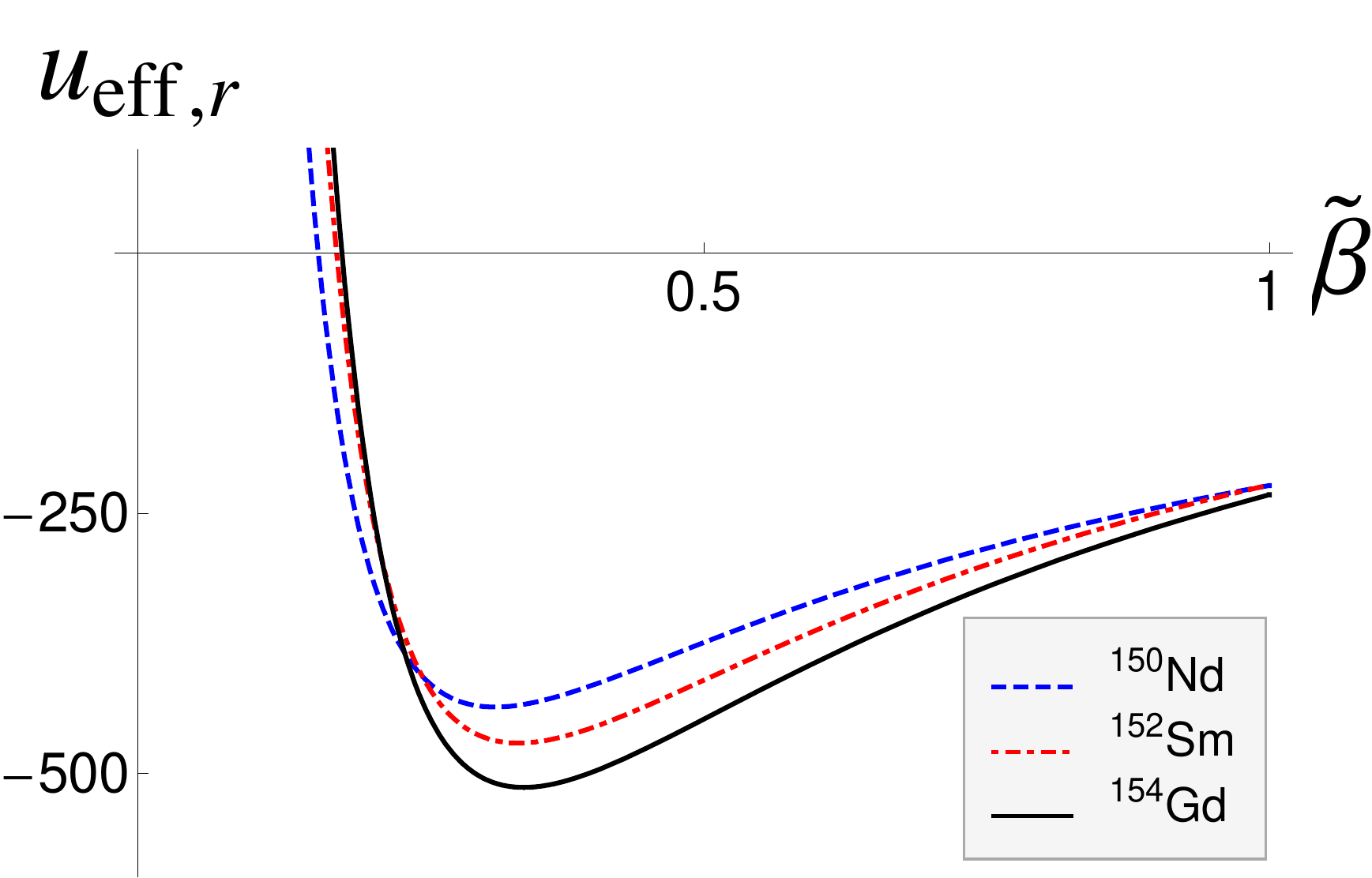}
\caption{(Color online) Effective potentials (Eq. (\ref{eq:ueff})) for $L=4$  for the ${\rm N}=90$ isotones being good examples 
of the X(5) critical point symmetry, corresponding to the parameters of Table II.  The quantities shown are dimensionless. Rescaling has been carried out, according to Appendix 7, 
using the rescaling parameters $A$ given in Table VIII. 
The rescaled effective potentials are given by Eq. (\ref{eff3}), while the rescaled abscissa, $\tilde \beta$, is given by Eq. (\ref{tildeb}). }
\end{figure}

\begin{figure*}[hbt]
\includegraphics[width=1\textwidth]{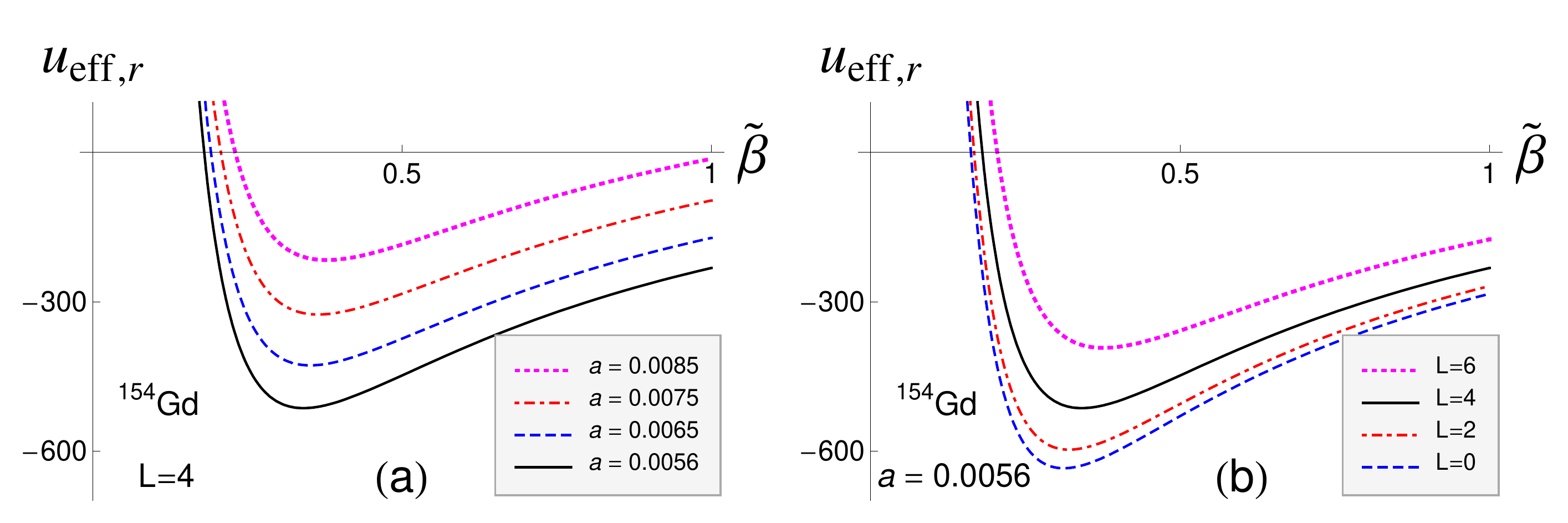}
\caption{(Color online) Dependence of the effective potentials of Eq. (\ref{eq:ueff}) on the parameter $a$ [panel (a)] 
and on the angular momentum $L$ [panel (b)]. The quantities shown are dimensionless. The effective potential for the $L=4$ state of the ground state band of $^{154}$Gd 
(corresponding to the parameters of Table II), plotted in both panels, is used as the basis of the comparison. 
Rescaling has been carried out, according to Appendix 7, using the rescaling parameters $A$ given in Table VIII. 
The rescaled effective potentials are given by Eq. (\ref{eff3}), while the rescaled abscissa, $\tilde \beta$, is given by Eq. (\ref{tildeb}). }
\end{figure*}

The dependence of the (rescaled according to Appendix 7) effective potentials of Eq. (\ref{eq:ueff}) on the parameter $a$ and on the angular momentum $L$ 
is shown in Fig. 6. The effective potential of the $L=4$ state of the ground state band of $^{154}$Gd is used as the basis
for the comparison. It is again seen that the effective potential becomes less deep as the parameter $a$ is increased. 
It also becomes less deep as the angular momentum $L$ is increased. 

\subsection{$B(E2)$s of $\gamma$-unstable nuclei}\label{BE2gam}

$B(E2)$s within the ground state band, as well as interband $B(E2)$s for which experimental 
data exist for several nuclei, have been calculated using the procedure described in subsec. X.A of Ref. \cite{DDMD}. 
For each nucleus, the parameters obtained by fitting the spectra have been used. 
The results are shown in Table V, the overall agreement being good.

\subsection{$B(E2)$s of axially symmetric deformed nuclei}\label{BE2def}

$B(E2)$s within the ground state band, as well as interband $B(E2)$s for which experimental 
data exist for several nuclei, have been calculated using the procedure described in subsec. X.B of Ref. \cite{DDMD}.
The results are shown in Table VI. In order to facilitate comparison with the ${\rm N}=90$ isotones $^{150}$Nd, $^{152}$Sm, 
$^{154}$Gd, and $^{156}$Dy, which are considered \cite{RMP82} as the best examples of the X(5) \cite{IacX5} critical point symmetry,
the X(5) predictions \cite{IacX5,BonX5,Bijker} are shown in the first line of the table.  
The overall agreement is good, although interband transitions are usually overestimated by theory. 

\section{Conclusions}

The main results of the present work are summarized here.

(i) Analytical solutions of the deformation dependent mass (DDM) Bohr Hamiltonian with the Kratzer potential have been obtained 
for $\gamma$-unstable, axially symmetric prolate deformed, and triaxial nuclei. The deformation function for the Kratzer potential 
was found to be $f(\beta)=1+a \beta$, to be compared with the deformation function $f(\beta)=1+a \beta^2 $ obtained 
in the case of the Davidson potential \cite{DDMD}.    

(ii) Despite the fact that the Davidson and Kratzer potentials have very different shapes,
numerical results coming from fitting the spectra of more than 100 nuclei and then using 
the same parameters for calculating $B(E2)$ transition rates, indicate that good overall agreement to 
the experimental data is achieved in both cases. Therefore the important factor in obtaining spectra 
with proper dependence of the moment of inertia on the deformation is the dependence of the mass
on the deformation, independently of the particular shape of the potential. 

(iii) On the other hand, a substantial difference between the two potentials shows up in the level spacings within the $\beta_1$ band.
While these spacings are usually overestimated by collective models based on the Bohr Hamiltonian, the DDM Davidson model included,
the present DDM Kratzer model avoids this problem, using the same number of parameters as the DDM Davidson model.  

(iv) Furthermore, the Kratzer potential is able to acquire the shape of a deep narrow well, while the Davidson potential cannot achieve this.
As a result, the DDM Bohr Hamiltonian with the Kratzer potential
succeeds in describing the ${\rm N}=90$ rare earths which are considered as hallmark examples \cite{McCutchan,RMP82} 
of the X(5) critical point symmetry \cite{IacX5} between spherical and deformed nuclei.     

The physical content of the free parameter $a$ appearing in the dependence of the mass on the deformation 
can be further investigated in at least two ways:

(i) By exploiting the equivalence of the deformation dependence of the mass to a curved space \cite{QT4267}. 

(ii) By considering the similarity of several terms appearing in the DDM Bohr Hamiltonian
to the Hamiltonian of the classical limit \cite{GK} of the interacting boson model \cite{IA}.

Work in these directions is in progress.   

\section*{Acknowledgements}

The authors are thankful to F. Iachello for suggesting the project and for useful discussions.
One of authors (N. M.) acknowledges the support of the Bulgarian Scientific Fund under contract DID-02/16-17.12.2009. 


\section*{Appendix 1: Deformed shape invariance} 

$H$ may be considered as the first member $H_0 = H$ of a hierarchy of Hamiltonians
\begin{equation}
  H_i = A_i^+ A_i^- + \sum_{j=0}^i \varepsilon_j, \qquad i=0, 1, 2, \ldots,
\end{equation}
where the first-order operators \cite{Q2929} 
\begin{equation}
  A_i^{\pm} = A^{\pm}(\mu_i, \nu_i) = \mp \sqrt{f} \frac{d}{d\beta} \sqrt{f} + W(\mu_i, 
  \nu_i; \beta) 
\end{equation}
satisfy a deformed shape invariance condition
\begin{equation}
  A_i^- A_i^+ = A_{i+1}^+ A_{i+1}^- + \varepsilon_{i+1}, \qquad i=0, 1, 2, \ldots,
\end{equation}
with $\varepsilon_i$, $i=0$, 1, 2,~\ldots, denoting some constants. 
In other words, the superpotential $W(\mu, \nu; \beta)$ fulfils the two conditions
\begin{equation}
  W^2(\mu, \nu; \beta) - f(\beta) W'(\mu, \nu; \beta) + \varepsilon_0 = 2 u_{\rm eff}(\beta), 
  \label{eq:SI-1}
\end{equation}
and
\begin{multline}
  W^2(\mu_i, \nu_i; \beta) + f(\beta) W'(\mu_i, \nu_i; \beta) = \\
  W^2(\mu_{i+1}, \nu_{i+1}; \beta) - f(\beta) W'(\mu_{i+1}, \nu_{i+1}; \beta) + \varepsilon_{i+1}, \\
  i=0, 1, 2, \ldots,  \label{eq:SI-2}
\end{multline}
where $\mu_0 = \mu$, $\nu_0 = \nu$, and a prime denotes derivative with respect to $\beta$.\par
%
%
In the case of the effective potential given in Eq.~(\ref{eq:ueff}), $W(\mu, \nu; \beta)$ is a class 1 superpotential
\begin{align}
  & W(\mu, \nu; \beta) = \mu \phi(\beta) + \nu, \label{eq:W}\\
  & \phi(\beta) = \frac{1}{\beta},  \label{eq:phi}
\end{align}
which means that Eqs.~(3.9) and (3.10) of \cite{Q2929} read
\begin{equation}
\begin{split}
  & \phi'(\beta) = - \frac{1}{\beta^2} = \frac{A}{\beta^2} + \frac{B}{\beta} + C, \\
  & a \beta = \frac{(A'/\beta^2) + (B'/\beta) + C'}{(-1/\beta^2)},
\end{split}
\end{equation}
with $A = -1$, $B = C = 0$, $A' = 0$, $B' = - a$ and $C' = 0$.\par
%
%
Inserting Eqs.~(\ref{eq:W}) and (\ref{eq:phi}) in (\ref{eq:SI-1}), we obtain
\begin{equation}
  \left(\frac{\mu}{\beta} + \nu \right)^2 - (1 + a \beta) \left(- \frac{\mu}{\beta^2}\right) +
   \varepsilon_0 = k_0 + \frac{k_{-1}}{\beta} + \frac{k_{-2}}{\beta^2}, 
\end{equation}
which is equivalent to the three equations
\begin{equation}
  \mu (\mu+1) = k_{-2}, \quad \mu (2\nu + a) = k_{-1}, \quad \nu^2 + \varepsilon_0 = k_0.
  \label{eq:3cond}
\end{equation}
Their solutions read
\begin{equation}\label{eq:Delta}
\begin{split}
  &\mu = \frac{1}{2} (- 1 \pm \Delta), \qquad \nu = \frac{k_{-1} - a\mu}{2\mu}, \qquad
         \varepsilon_0 = k_0 - \nu^2, \\
  & \Delta \equiv \sqrt{1 + 4 k_{-2}}, 
\end{split}  
\end{equation}
where we note that $1 + 4 k_{-2}$ is always positive. As we shall show in Appendix 2, the conditions ensuring that the ground state wavefunction is physically acceptable select the lower sign for $\mu$:
\begin{equation}
  \mu = - \frac{1}{2} (1 +\Delta).  
  \label{eq:lambda-mu}
\end{equation}
\par
%
%
Inserting next Eqs.~(\ref{eq:W}) and (\ref{eq:phi}) in Eq.~(\ref{eq:SI-2}), we get
\begin{multline}
  \left(\frac{\mu_i}{\beta} + \nu_i\right)^2 + (1 + a \beta) \left(- \frac{\mu_i}{\beta^2}\right) = \\ 
  \left(\frac{\mu_{i+1}}{\beta} + \nu_{i+1}\right)^2 - (1 + a \beta) \left(- \frac{\mu_{i+1}}
  {\beta^2}\right) + \varepsilon_{i+1},
\end{multline}
leading to the three conditions
\begin{equation}
\begin{split}
  & \mu_i (\mu_i - 1) = \mu_{i+1} (\mu_{i+1} + 1), \\ 
  & \mu_i (2\nu_i - a) = \mu_{i+1} (2\nu_{i+1} + a), \\
  & \nu_i^2 = \nu_{i+1}^2 + \varepsilon_{i+1}.  
\end{split}
\end{equation}
Their solutions are
\begin{multline}
  \mu_{i+1} = \mu_i - 1, \qquad 2 \mu_{i+1} \nu_{i+1} = 2 \mu_i \nu_i - a (\mu_i + \mu_{i+1}), \\
  \varepsilon_{i+1} = \nu_i^2 - \nu_{i+1}^2.  \label{eq:sol-SI-2}
\end{multline}
Note that there is an other solution for $\mu_{i+1}$, namely $\mu_{i+1} = - \mu_i$, but the alternating signs would not be compatible with physically acceptable excited state wave functions. Finally, the iteration of the first two relations in (\ref{eq:sol-SI-2}) leads to
\begin{equation}
  \mu_i = \mu - i, \qquad \nu_i = \frac{2\mu (\nu - a i) + a i^2}{2(\mu - i)}.  \label{eq:lambda-mu-i}
\end{equation}
\par
%
%
{}For future use, it is worth observing that $\nu_i$ can be rewritten in terms of $\mu_i$ as
\begin{equation}
  \nu_i = \frac{a}{2} \left(\mu_i - \frac{\tK}{\mu_i}\right), \qquad \tK \equiv k_{-2} - \frac{k_{-1}}{a}.
  \label{eq:K}
\end{equation}
From (\ref{eq:lambda-mu-i}), it indeed follows that 
\begin{equation}
  \nu_i = \frac{a}{2} \left(\mu_i - \frac{\mu \left(\mu - \frac{2\nu}{a}\right)}{\mu_i}\right),  
\end{equation}
while Eq. (\ref{eq:3cond}) leads to
\begin{equation}
  k_{-2} - \frac{k_{-1}}{a} = \mu \left(\mu - \frac{2\nu}{a}\right).
\end{equation}
The energy spectrum of Eq.~(\ref{eq:e14}) is therefore given by
\begin{multline}
  \epsilon_n  = \tfrac{1}{2} \sum_{i=0}^n \varepsilon_i = \tfrac{1}{2}\left(k_0 - \nu_n^2\right) \\
   = \frac{1}{2} \left[k_0 - \left(\frac{k_{-1} + a \left[n^2 + \frac{1}{2}(1 + \Delta)(2n+1)\right]}
       {2n+1+\Delta}\right)^2\right],\\ 
     n=0, 1, 2, \ldots.
\end{multline}
%
\section*{Appendix 2: Conditions for physically acceptable wave functions}

To be physically acceptable, the bound-state wavefunctions should satisfy two conditions \cite{Q2929}:\par
\noindent (i) As in conventional (constant-mass) quantum mechanics, they should be square integrable on
the interval of definition of $u_{\rm eff}$, i.e.,
\begin{equation}
  \int_{0}^{\infty} d\beta\,  |R_n(\beta)|^2 < \infty.  \label{eq:wf-C1}
\end{equation}
\noindent (ii) Furthermore, they should ensure the Hermiticity of $H$. For such a purpose, it is enough to impose that the operator $- {\rm i}\sqrt{f} (d/d\beta) \sqrt{f}$ be Hermitian, which amounts to the restriction
\begin{equation}
  |R_n(\beta)|^2 f(\beta) \to 0 \qquad {\rm for\ } \beta \to 0 {\rm \ and\ } \beta \to \infty,
\end{equation}
or, equivalently,
\begin{multline}
  |R_n(\beta)|^2 \to 0 \quad {\rm for\ } \beta \to 0 \qquad {\rm and} \\
  \qquad |R_n(\beta)|^2 \beta 
  \to 0 \quad {\rm for\ } \beta \to \infty.  \label{eq:wf-C2}
\end{multline}
As condition (\ref{eq:wf-C2}) is more stringent than condition (\ref{eq:wf-C1}), we should only be concerned with the former.\par
%
%
\section*{Appendix 3: Ground-state wavefunction} \label{gswf}

The ground-state wavefunction, which is annihilated by $A^-$, is given by Eq.~(2.29) of \cite{Q2929} as
\begin{multline}
  R_0(\beta) = R_0(\mu, \nu; \beta) \\
  = \frac{N_0}{\sqrt{f(\beta)}} \exp \left(- \int^{\beta}
  \frac{W(\mu, \nu; \tbeta)}{f(\tbeta)} d\tbeta\right),
\end{multline}
where $N_0$ is some normalization coefficient. Here
\begin{multline}
  \int^{\beta} \frac{W(\mu, \nu; \tbeta)}{f(\tbeta)} d\tbeta = \int^{\beta} \left(\frac{\mu}
  {\tbeta} + \frac{\nu - a\mu}{1 + a \tbeta}\right) d\tbeta \\
  = \mu \ln \beta + \frac{1}{a} (\nu - a\mu) \ln (1 + a \beta).
\end{multline}
Hence
\begin{equation}
  R_0(\beta) = N_0 \beta^{-\mu} f^{- \frac{\nu - a \mu}{a} - \frac{1}{2}} = N_0 \beta^{-\mu} 
  f^{\frac{1}{2} \left(\mu + \frac{\tK}{\mu} - 1\right)},  \label{eq:R_0}
\end{equation}
where in the last step we used Eq.\ (\ref{eq:K}) for $i=0$.\par
%
%
{}For $\beta \to 0$, the function $|R_0(\beta)|^2$ behaves as $\beta^{-2\mu}$. Condition (\ref{eq:wf-C2}) imposes that $\mu < 0$. Since $k_{-2}$, defined in Eq.~(\ref{eq:ueff}), is greater than 2, it follows that $\Delta$, defined in (\ref{eq:Delta}), is greater than 3, so that the upper sign choice for $\mu$ in (\ref{eq:Delta}) would lead to $\mu > 1$. As this is not acceptable, we have to take the lower sign for which $\mu < -2$.\par
%
%
{}For $\beta \to \infty$, $|R_0(\beta)|^2 \beta$ behaves as $\beta^{- \mu + \frac{\tK}{\mu}}$. Condition (\ref{eq:wf-C2}) therefore imposes that $\tK /\mu < \mu$ or
\begin{equation}
  \mu^2 < \tK.
\end{equation}
On taking (\ref{eq:3cond}), (\ref{eq:lambda-mu}) and (\ref{eq:K}) into account, this restriction can be rewritten as
\begin{equation}
  |\mu| = \frac{1}{2} (1+\Delta) < - \frac{k_{-1}}{a}.  \label{eq:cond-gs}
\end{equation}
A physically acceptable ground-state wavefunction therefore exists only if the inequality (\ref{eq:cond-gs}) is satisfied by the parameters of the problem.\par
%
%
\section*{Appendix 4: Excited-state wavefunctions}

According to Eqs.~(2.30), (3.18) and (3.19) of \cite{Q2929}, the excited-state wavefunctions are given by
\begin{multline}
  R_n(\beta) = R_n(\mu, \nu; \beta) \propto R_0(\mu_n, \nu_n; \beta) P_n(\mu, \nu; y), \\
   y = \frac{1}{\beta},  \label{eq:R_n}
\end{multline}
where $P_n(\mu, \nu; y)$ is an $n$th-degree polynomial in $y$, satisfying the equation
\begin{multline}
  P_{n+1}(\mu, \nu; y) = y (y+a) \frac{d}{dy} P_n(\mu_1, \nu_1; y) \\
  + [(\mu_{n+1} + \mu) y + \nu_{n+1} + \nu ] P_n(\mu_1, \nu_1; y),  \label{eq:P-eq}
\end{multline}
with the starting value $P_0(\mu, \nu; y) = 1$.\par
%
%
{}From Eqs.~(\ref{eq:K}) and (\ref{eq:R_0}), it follows that 
\begin{multline}
  R_n(\beta) \propto \beta^{-\mu_{n}} f^{- \frac{\nu_n - a \mu_n}{a} - \frac{1}{2}} P_n(\mu, \nu; y) \\
  \propto \beta^{-\mu_{n}} f^{\frac{1}{2} \left(\mu_n + \frac{\tK}{\mu_n} - 1\right)} P_n(\mu, \nu; y).
  \end{multline}
\par
%
%
{}For $\beta \to 0$, the function $|R_n(\beta)|^2$ behaves as $\beta^{- 2\mu_n - 2n} = \beta^{-2\mu}$, which goes to zero 	as it should be. For $\beta \to \infty$, $|R_n(\beta)|^2 \beta$ behaves as $\beta^{- \mu_n + \frac{\tK}{\mu_n}}$. Condition (\ref{eq:wf-C2}) therefore imposes that 
\begin{equation}
  \mu_n^2 < \tK,
\end{equation}
which is equivalent to
\begin{equation}
  n^2 + (2n+1) |\mu| = n^2 + \left(n + \frac{1}{2}\right) (1+\Delta) < - \frac{k_{-1}}{a}.  
\end{equation}
We conclude that there is a finite number of bound-state wavefunctions corresponding to $n=0$, 1,~\ldots, $n_{\rm max}$, where $n_{\rm max}$ is such that
\begin{multline}
  n_{\rm max}^2 + \left(n_{\rm max} + \frac{1}{2}\right) (1+\Delta) < - \frac{k_{-1}}{a} \\ 
  \le (n_{\rm max} + 1)^2 + \left(n_{\rm max} + \frac{3}{2}\right) (1+\Delta).
\end{multline}
\par
%
%
It now remains to solve Eq.~(\ref{eq:P-eq}). For such a purpose, let us make the changes of variable and of function
\begin{equation}
  t = \frac{2y + a}{a}, \quad P_n\left(\mu, \frac{a}{2}\left(\mu - \frac{\tK}{\mu}\right); y\right) = 
  C_n^{(\mu)} Q_n(\mu; t),  \label{eq:change}
\end{equation}
where $C_n^{(\mu)}$ is some constant. From definition (\ref{eq:change}), it follows that $Q_n(\mu; t)$ is an $n$th-degree polynomial in $t$. We successively get
\begin{equation}
  y = \frac{a}{2} (t-1), \qquad y + a = \frac{a}{2} (t+1), \qquad \frac{d}{dy} = \frac{2}{a} \frac{d}{dt}.
\end{equation}
By taking Eq.~(\ref{eq:K}) into account, it is then straightforward to show that Eq.~(\ref{eq:P-eq}) becomes
\begin{equation}
\begin{split}
  \frac{2 C_{n+1}^{(\mu)}}{a C_n^{(\mu-1)}} Q_{n+1}(\mu; t) = \left\{ - (1 - t^2) \frac{d}{dt} + 
  (2 \mu_{n+1} + n + 1) \right. \\
 \left.  \times \left(t - \frac{\tK}{\mu_{n+1} (\mu_{n+1} + n + 1)}\right)\right\} \quad Q_n(\mu - 1; t).
  \end{split}
\end{equation}
\par
%
%
On taking into account that the Jacobi polynomials satisfy the equation
\begin{multline}
  \frac{2(n+1)(n+\alpha+\beta+1)}{2n+\alpha+\beta+2} P_{n+1}^{(\alpha, \beta)} (z) \\
  = \left\{- (1 - z^2) \frac{d}{dz} + (n + \alpha + \beta + 1) \right. \\
 \left. \times   \left(z + \frac{\alpha-\beta}{2n+\alpha+
        \beta+2}\right)\right\} P_n^{(\alpha, \beta)}(z),
\end{multline}
obtained by combining their recurrence and differential relations (Eqs.~(22.7.1) and (22.8.1) of \cite{AbrSte}),
we see that $Q_n(\mu-1; t)$ is actually some Jacobi polynomial
\begin{equation}
  Q_n(\mu-1; t) = P_n^{\left(\mu_{n+1} - \frac{\tK}{\mu_{n+1}}, \mu_{n+1} + \frac{\tK}{\mu_{n+1}}
  \right)}(t),
\end{equation}
provided we choose
\begin{equation}
  \frac{C_{n+1}^{(\mu)}}{C_n^{(\mu-1)}} = \frac{a}{2} \frac{(n+1) (2\mu_{n+1}+n+1)}
  {\mu_{n+1}+n+1}, \quad C_0^{(\mu-n-1)} = 1,
\end{equation}
or, in other words, 
\begin{equation}
  C_n^{(\mu)} = \left(\frac{a}{2}\right)^n \ \frac{n! \Gamma(2\mu-n+1) \Gamma(\mu-n+1)}
  {\Gamma(2\mu-2n+1) \Gamma(\mu+1)}.
\end{equation}
%
%
We therefore conclude that the wavefunctions are given by
\begin{multline}
  R_n(\beta) = N_n \beta^{- \mu_n} f^{\frac{1}{2} \left(\mu_n + \frac{\tK}{\mu_n} - 1\right)} 
  P_n^{\left(\mu_n - \frac{\tK}{\mu_n}, \mu_n + \frac{\tK}{\mu_n}\right)}(t), \\
  t = \frac{2+a\beta}{a\beta},
\end{multline}
where $N_n$ is some normalization coefficient.\par
%
%
\section*{Appendix 5: Normalization coefficient}

To calculate $N_n$, let us first express the whole wavefunction $R_n$ in terms of $t$:
\begin{multline}
  R_n = N_n \left(\frac{a}{2}\right)^{\mu_n} (t-1)^{\frac{1}{2}\left(\mu_n - \frac{\tK}{\mu_n} + 1
  \right)} \\
  (t+1)^{\frac{1}{2}\left(\mu_n + \frac{\tK}{\mu_n} - 1\right)} P_n^{\left(\mu_n - \frac{\tK}
  {\mu_n}, \mu_n + \frac{\tK}{\mu_n}\right)}(t). 
\end{multline}
From this we get
\begin{multline}
  \int_0^{\infty} |R_n(\beta)|^2 d\beta = |N_n|^2 \left(\frac{a}{2}\right)^{2\mu_n - 1} \int_1^{\infty}
  (t-1)^{\mu_n - \frac{\tK}{\mu_n} - 1} \\
\times  (t+1)^{\mu_n + \frac{\tK}{\mu_n} - 1} \left[P_n^{\left(\mu_n - 
  \frac{\tK}{\mu_n}, \mu_n + \frac{\tK}{\mu_n}\right)}(t)\right]^2. 
 \end{multline}
\par
%
%
{}For $n=0$, the integral is easily calculated by using Eq.~(3.196.2) of \cite{Grad}. This leads to
\begin{equation}\label{Nzero}
  N_0 = a^{- \mu + \frac{1}{2}} \left(\frac{\Gamma\left(- \mu - \frac{\tK}{\mu} + 1\right)}{\Gamma
  (- 2\mu + 1) \Gamma\left(\mu - \frac{\tK}{\mu}\right)}\right)^{1/2}. 
\end{equation}
For $n \ne 0$, a similar integral follows from expanding the two Jacobi polynomials according to Eq.~(22.3.1) 
of \cite{AbrSte} and integrating term by term. The result for $N_n$ reads
\begin{multline}\label{Nall}
  N_n = a^{- \mu_n + \frac{1}{2}} \\
   \times \left[\Gamma\left(\mu_n - \frac{\tK}{\mu_n} + n + 1\right)
        \Gamma\left(\mu_n + \frac{\tK}{\mu_n} + n + 1\right)\right]^{-1} \\
  \times \left(\Gamma(- 2\mu + 1) \sum_{m,p=0}^n \frac{A_n^{(m,p)}}{B_n^{(m,p)}}\right)
        ^{-1/2},
\end{multline}
where
\begin{equation}
\begin{split}
  A_n^{(m,p)} = \Gamma\left(\mu_n - \frac{\tK}{\mu_n} + 2n - m - p\right), \\ 
  B_n^{(m,p)} = m!\, (n-m)!\, p!\, (n-p)!\, \\ 
  \times \Gamma\left(\mu_n - \frac{\tK}{\mu_n} + n - m + 1\right)
       \Gamma\left(\mu_n + \frac{\tK}{\mu_n} + m + 1\right) \\
  \times \Gamma\left(\mu_n - \frac{\tK}{\mu_n} + n - p + 1\right) \Gamma\left(\mu_n + 
        \frac{\tK}{\mu_n} + p + 1\right) \\
 \times \Gamma\left(- \mu_n - \frac{\tK}{\mu_n} - m - p + 1\right).
\end{split}
\end{equation}
\par

\section*{Appendix 6: Wave functions in the $a\to 0$ limit} 

In the present DDM Kratzer solution the wave functions contain Jacobi polynomials, 
while in the usual case \cite{FV1,FV2} they contain Laguerre polynomials. 
Contact between the two results should be established.

In the limit of no dependence of the mass on the deformation, i.e., $a\to 0$, one has
\begin{equation}
k_0=0, \qquad k_{-1}=-2,
\end{equation}
as well as  
\begin{equation}\label{Klimits}
 \bar K = k_{-2}+{2\over a}, \qquad t=1+{2\over a \beta}.
 \end{equation}
The last two quantities become infinite in the $a\to 0$ limit. 
 
One should first notice that in the Jacobi polynomial of the present solution 
the variable $\beta$ appears in the denominator, since $t=1+{2 \over a \beta}$,
while 
in the Laguerre polynomials of the usual solution it appears in the numerator.
Thus one should first use the identity (22.5.43 of \cite{AbrSte})
\begin{multline}
P_n^{(\alpha,\beta)}(x)={2n+\alpha+\beta \choose n} \left({x-1\over 2} \right)^n \\
{}_2F_1\left(-n,-n-\alpha ; -2n-\alpha-\beta; {2\over 1-x}\right),
\end{multline}
which in the present case leads to
\begin{multline}\label{2F1}
P_n^{\left(\mu_n - \frac{\bar K}{\mu_n}, \mu_n + \frac{\bar K}{\mu_n}\right)}(t)=
{2\mu \choose n} {1\over (a\beta)^n} \quad \\
{}_2F_1\left(-n, {\bar K\over \mu-n}-\mu ; -2\mu ; -a\beta\right).
\end{multline}

One should now implement the connection between hypergeometric functions and confluent hypergeometric functions. 
The hypergeometric functions ${}_2F_1(a,b;c;z)$ are solutions of the equation (p. 146 of Ref. \cite{Greiner})
\begin{equation}
z(1-z) {d^2\phi \over dz^2} +[c-(a+b+1)z] {d\phi\over dz}-ab\phi=0.
\end{equation}
Using the linear transformation $x=bz$ this leads to (page 148 of Ref. \cite{Greiner})
\begin{equation}
x\left(1-{x\over b}\right) {d^2\phi \over dx^2} +\left[c-(a+1){x\over b}-x\right] {d\phi\over dx}-a\phi=0.
\end{equation}
In the limit $b\to \infty$ this gives the equation
\begin{equation}
x {d^2\phi \over dx^2} +(c-x) {d\phi\over dx}-a\phi=0,
\end{equation}
which has as solutions the confluent hypergeometric functions ${}_1F_1(a;c;x)$. 
As a result, in the limit $b\to \infty$, the hypergeometric function ${}_2F_1(a,b;c;z)$
is reduced to the confluent hypergeometric function ${}_1F_1(a;c;x)$, where  $x=bz$.

In the present case we consider the hypergeometric function  
\begin{equation}
{}_2F_1\left(-n, {\bar K\over \mu-n}-\mu ; -2\mu ; -a\beta\right).
\end{equation}
As mentioned above, in the limit $a\to 0$ the term $\bar K$ becomes infinite.
Thus the condition for the transition from ${}_2F_1(a,b;c;z)$ to ${}_1F_1(a;c;x)$
applies, leading to 
\begin{multline}
{}_1F_1\left(-n ; -2\mu ; -\left( {\bar K\over \mu-n} -\mu\right)a\beta\right) \\
={}_1F_1\left(-n ; \Delta+1 ; -\left( {\bar K\over \mu-n} -\mu\right)a\beta\right).
\end{multline}
One can now use the relation between confluent hypergeometric functions and Laguerre polynomials 
(p. 149 of Ref. \cite{Greiner})
\begin{equation}\label{FL}
{}_1F_1(-n; m+1; z) = {n! m!\over (n+m)!} L_n^{(m)}(z),
\end{equation}
leading in the present case to
\begin{multline}\label{1F1}
{}_1F_1\left(-n ; \Delta+1 ; -\left( {\bar K\over \mu-n} -\mu\right)a\beta\right) \\
={n! \Delta! \over (n+\Delta)!} L_n^{(\Delta)} \left( -\left( {\bar K\over \mu-n} -\mu\right)a\beta\right).
\end{multline}
From what we have seen above, $n$ and $\mu = - \frac{1}{2} (1 +\Delta)$ remain finite, 
while $\bar K = k_{-2}+{2\over a}$ becomes infinite because of the $1/a$ term, with $k_{-2}$ 
remaining finite. Therefore in the argument of the Laguerre polynomial we can ignore the $\mu$ term 
as much smaller than the term containing $\bar K$, while within $\bar K$ we can ignore the finite term $k_{-2}$
in comparison to $1/a$, getting
\begin{multline}\label{ourF}
{}_1F_1 \left( -n; \Delta+1; {2\beta \over {1\over 2}(1+\Delta)+n} \right) \\
= {n! \Delta! \over (n+\Delta)!} L_n^{(\Delta)} \left( {2\beta \over {1\over 2}(1+\Delta)+n} \right).
\end{multline}

Concerning the factor  $f^{\frac{1}{2} \left(\mu_n + \frac{\bar K}{\mu_n} - 1\right)} $ in Eq. (\ref{RRn}),
we remark that in the $a\to 0$ limit the term $\bar K$ goes to infinity, while the other terms do not.
From Eq. (\ref{Klimits}) it is clear that in this limit $\bar K$ can be replaced by $2/a$. 
Thus one can write 
\begin{equation}
F = f^{\frac{1}{2} \left(\mu_n + \frac{\bar K}{\mu_n} - 1\right)} \to f^{1\over a(\mu -n)} = 
\left( \left( 1+ {\beta \over a^{-1}} \right)^{a^{-1}}\right)^{1\over \mu-n}.
\end{equation}
Taking advantage of Eq. (4.2.21) of Ref. \cite{AbrSte}
\begin{equation}
\lim _{m\to \infty} \left( 1+ {z\over m}\right)^m = e^z, 
\end{equation}
the above factor becomes 
\begin{equation}\label{FF}
F = \left( e^\beta \right)^{1\over \mu-n} = \exp\left(-{\beta \over n+{1\over 2} +{\Delta \over 2}}\right).
\end{equation}

\subsection*{6.1 $\gamma$-unstable nuclei}

The wave functions for $\gamma$-unstable nuclei are given in Ref. \cite{FV1} (not normalized) and in Ref. \cite{Wolf}
(normalized). Using the notation of Ref. \cite{FV1} the normalized wave functions take the form 
\begin{equation}\label{FVunstable}
\chi(x)_{\tau,n}=\sqrt{n!\over \Gamma(n+2\mu+1)} x^{\mu+{1\over 2}} e^{-x/2} L_n^{(2\mu)}(x),
\end{equation}
where 
\begin{multline}
x=2\beta \sqrt{\epsilon}, \qquad \mu= \sqrt{\left(\tau+{3\over 2} \right)^2+ D \beta_0^2},\\  
\sqrt{\epsilon}={D \beta_0 \over \mu +n+{1\over 2} },
\end{multline}
and taking into account that in Ref. \cite{FV1}, the power of $x$
appears as $2\mu+1$ instead of $\mu+{1\over 2}$, because of a misprint.

In the DDM Kratzer case one has 
\begin{equation}
\Lambda = \tau (\tau
+3)=\frac{L}{2}\left(\frac{L}{2}+3\right)=\frac{1}{4}L(L+6),
\end{equation}
thus one gets 
\begin{multline}
k_{-2}=2 + \tau(\tau+3)+2\tilde B, \\
 \Delta=2 \sqrt{ \left(\tau+{3\over 2} \right)^2+2\tilde B}, 
\end{multline}
leading in Eq. (\ref{FVunstable}) the Laguerre polynomial into the form 
\begin{equation}\label{ourL}
 L_n^{\left(2 \sqrt{ \left(\tau+{3\over 2} \right)^2+2\tilde B}\right)}  \left( {2\beta \over \sqrt{ \left(\tau+{3\over 2} \right)^2+2\tilde B}+{1\over 2}+n} \right).
\end{equation}
We see that with the parameter correspondence between the DDM Kratzer case and Ref. \cite{FV1}
\begin{equation}
2\tilde B\to \beta_0, \qquad 1\to  D \beta_0.
\end{equation}
the Laguerre polynomials in the two expressions are identical. 
The exponential factor in Eq. (\ref{FVunstable}) is also identical to the one of Eq. (\ref{FF}).  

Written in the DDM parametrization the wave function of Ref. \cite{FV1} reads  
\begin{multline}\label{Vitturi}
\chi(\beta)_{\tau,n}=\sqrt{n!\over \Gamma(n+2\mu+1)} 
\left({2\beta\over \mu+{1\over 2}+n} \right)^{\mu+{1\over 2}}  \\
\exp\left({-\beta \over \mu +{1\over 2}+ n }\right) L_n^{(2\mu)}\left({2\beta \over \mu +{1\over 2}+ n }\right),
\end{multline}
where 
\begin{equation}
\mu = \sqrt{\left(\tau+{3\over 2}  \right)^2 +2 \tilde B}.
\end{equation}

\subsection*{6.2 Axially symmetric prolate deformed nuclei} 

The wave functions for axially symmetric prolate deformed nuclei are given in Ref. \cite{FV2} (not normalized) and in Ref. \cite{Wolf}
(normalized). Using the notation of Ref. \cite{FV2} the normalized wave functions take again the form of Eq. (\ref{FVunstable}), but with 
\begin{multline}
x=2\beta \sqrt{\epsilon}, \qquad \mu= \sqrt{  {L(L+1)\over 3}+{9\over 4}+ B},\\  
\sqrt{\epsilon}={A/2 \over \mu +n+{1\over 2} }.
\end{multline}

In the DDM Kratzer case one has 
\begin{equation}
\Lambda = \frac{L(L+1)-K^2}{3} + 6c(n_{\gamma}+1),  \ \ \ \ c=\frac{C}{2}. 
\end{equation}
Since in Ref. \cite{FV2} the approximate separation of variables introduced in X(5) is used,
we formally put $c=0$. For the ground state band and the $\beta$-bands one also has $K=0$. 
As a result we get 
\begin{multline}
k_{-2}=2+ {L(L+1)\over 3} +2 \tilde B, \\ \Delta=2\sqrt{{9\over 4}+ {L(L+1)\over 3}+2 \tilde B },
\end{multline}
leading in Eq. (\ref{FVunstable}) the Laguerre polynomial into the form 
\begin{equation}
 L_n^{\left(2\sqrt{ {9\over 4}+ {L(L+1)\over 3}+2\tilde B}\right) } \left({2\beta \over \sqrt{ {9\over 4}+{L(L+1)\over 3}+2\tilde B}
+{1\over 2}+n }  
  \right).
\end{equation}
We see that with the parameter correspondence between the DDM Kratzer case and Ref. \cite{FV2}
\begin{equation}
2\tilde B\to B, \qquad 2\to  A,
\end{equation}
the Laguerre polynomials in the two expressions are identical. 
The exponential factor in Eq. (\ref{FVunstable}) is also identical to the one of Eq. (\ref{FF}).  

Written in the DDM parametrization the wave function of Ref. \cite{FV2} reads
again as Eq. (\ref{Vitturi}), but with   
\begin{equation}
\mu= \sqrt{  {L(L+1)\over 3}+{9\over 4}+ 2 \tilde B}.
\end{equation}

\section*{Appendix 7: Scaling factors}

In the present work the Kratzer potential has been used in the form of Eq. (\ref{Kratz}),
in which no free parameter appears in its first term. This choice does not affect ratios 
of energies and ratios of $B(E2)$s, as it will be confirmed below. In this way the addition 
of an extra free parameter is avoided. If, however, one wishes to plot the effective potentials
for different nuclei, this factor comes in and should be determined for each nucleus separately,
as we shall see below. 

\subsection*{7.1 Spectra} 

In the present work the Kratzer potential is used in the form of Eq. (\ref{Kratz}). 
In order to compare the results to the case in which a free parameter 
appears in the first term, we can use the rescaling transformation 
\begin{equation}\label{tildeb}
\beta = {\tilde \beta \over A}, 
\end{equation} 
leading to 
\begin{equation}\label{Br}
  u(\tilde \beta) = - \frac{A}{\tilde \beta} + \frac{\tilde{B_r} }{\tilde \beta^2}, \quad \tilde B_r= A^2 \tilde B .
\end{equation}
In what follows we are going to use the subscript $r$ for labelling rescaled quantities. 

Using this rescaling in the original Bohr Hamiltonian of Eq. (\ref{eq:e1}) we obtain 
\begin{equation} 
\begin{split}
 & H_B = -{\hbar^2 \over 2B_r} \left[ {1\over \tilde \beta^4} {\partial \over \partial 
\tilde \beta} \tilde \beta^4 {\partial \over \partial \tilde \beta} + {1\over \tilde \beta^2 \sin 
3\gamma} {\partial \over \partial \gamma} \sin 3 \gamma {\partial \over 
\partial \gamma} \right.  \\
& \left. - {1\over 4 \tilde \beta^2} \sum_{k=1,2,3} {Q_k^2 \over \sin^2 
\left(\gamma - {2\over 3} \pi k\right) } \right] +V(\beta,\gamma),
\end{split}
\end{equation}
with
\begin{equation}
B_r={B\over A^2}. \label{Hr} 
\end{equation}

When allowing the nuclear mass to depend on the deformation as in Eq. (\ref{massdep}), 
the rescaling leads to 
\begin{equation}
B_r(\tilde \beta)=\frac{{B_0}_r}{(f(\tilde \beta))^2}, \quad {B_0}_r = {B_0\over A^2},  
\end{equation}
the deformation function of Eq. (\ref{deffun}) becoming 
\begin{equation}\label{deffun2}
f(\tilde \beta)= 1 + a_r \tilde \beta, \quad a_r= {a \over A}.
\end{equation}

From Eq. (\ref{Hr}) it is clear that reduced energies, formerly defined through Eq. (\ref{reducedE}), now become 
\begin{equation}
\epsilon_r = {B_0}_r E/\hbar^2 = {\epsilon \over A^2}.  
\end{equation}
This implies that 
\begin{equation}\label{Escale}
\epsilon = A^2 \epsilon_r, 
\end{equation}
in agreement to Eq. (11) of Ref. \cite{FV1} and to Eq. (10) of Ref. \cite{FV2},
indicating that the energy levels scale as $A^2$. 

In the same way reduced potentials, formerly defined through Eq. (\ref{reducedV}), now become 
\begin{equation}\label{rescV}
v_r = {B_0}_r V/\hbar^2 = {v \over A^2},  
\end{equation}
implying
\begin{equation}\label{Vscale}
v = A^2 v_r. 
\end{equation}

In axially symmetric prolate deformed nuclei, in which the potential 
is given in Eqs. (\ref{eq:e7b}) and (\ref{wgamma}),  one gets 
\begin{equation}
v(\tilde \beta,\gamma)= u(\tilde \beta)+{f^2 \over \tilde \beta^2} w_r(\gamma), \quad w_r(\gamma)=A^2 w(\gamma)
\end{equation}
with 
\begin{equation}
w_r(\gamma)= {1\over 2} (3c_r)^2 \gamma^2 , \quad c_r = A c . 
\end{equation}

Similarly in triaxial nuclei with $\gamma \approx \pi/6$, in which the potential is assumed to be 
of the form of Eqs. (\ref{eq:e7b}) and (\ref{wgammatriax}), we  obtain 
\begin{equation}
w_r(\gamma)= {1\over 4} c_r \left( \gamma -{\pi\over 6}\right)^2  , \quad c_r= A^2 c.
\end{equation}

The effective potential of Eq. (\ref{eq:e15}), in the Kratzer case, in which $f(\beta)$ is given by Eq. (\ref{deffun}) and 
\begin{equation}
\quad f'=a, \quad f''=0, 
\end{equation}
 takes the form
\begin{eqnarray}\label{eff1}
u_{eff}= -{1\over \beta} + {\tilde B \over \beta^2} + (1-\delta-\lambda)  f { a\over \beta} 
 \nonumber \\ 
+ {1\over 2} \left( {1\over 2}-\delta\right) 
\left( {1\over 2}-\lambda\right) a^2 \nonumber \\
+ {f^2+a \beta f \over \beta^2}+ {f^2 \over 2 \beta^2}  \Lambda,    
\end{eqnarray}
with $f$ given by Eq. (\ref{deffun}). 
In rescaled notation the effective potential is written as 
\begin{eqnarray}
u_{eff}= -{A\over \tilde \beta} + {\tilde B_r \over \tilde \beta^2} + (1-\delta-\lambda)  f { a_r\over \tilde \beta} A^2  
 \nonumber \\ 
+ {1\over 2} \left( {1\over 2}-\delta\right) 
\left( {1\over 2}-\lambda\right) a_r^2 A^2 \nonumber \\
+ {f^2+a_r \tilde \beta f \over \tilde \beta^2}A^2 
+ {f^2 \over 2 \tilde \beta^2} A^2   \Lambda, \label{eff2}
\end{eqnarray}
with $f$ given by Eq. (\ref{deffun2}).  
The rescaled effective potential, $u_{eff,r}(\tilde \beta)$, is then calculated by plugging this result in the rhs of Eq. (\ref{rescV})
\begin{equation}\label{eff3}
u_{eff,r}={u_{eff}\over A^2}. 
\end{equation}

\subsection*{7.2 Position of the minimum of the effective potential}

It is instructive to study the position, $\beta_0$, of the minimum of the effective potential, given in Eq. (\ref{eq:ueff}). 
Equating the derivative of the effective potential to zero we easily find for $\delta=\lambda=0$
\begin{equation}\label{beta0}
\beta_0= -2{k_{-2}\over k_{-1}} = {2+\Lambda +2\tilde B \over 1 -4a-a\Lambda}.
\end{equation}

In the case of $\gamma$-unstable nuclei for the ground state ($\tau=0$, $L=0$) we obtain 
\begin{equation}\label{b0unst}
\beta_0= {2 + 2\tilde B \over 1-4 a}. 
\end{equation}
Using Eq. (\ref{tildeb}) this leads to 
\begin{equation}\label{Aunst}
A = {1-4a \over 2 + 2\tilde B} \tilde \beta_0.
\end{equation}

In the case of prolate deformed nuclei for the ground state ($L=0$, $n_\gamma=0$) we obtain 
\begin{equation}\label{b0prol}
\beta_0= {2 + 6c + 2\tilde B \over 1-4 a- 6 a c}. 
\end{equation}
Using Eq. (\ref{tildeb}) this leads to 
\begin{equation}\label{Aprol}
A = {1-4a-6 a c \over 2 + 6 c + 2\tilde B} \tilde \beta_0.
\end{equation}

\subsection*{7.3 Depth of the effective potential} 

As depth of the effective potential its value at the position of its minimum is considered. 
Substituting Eq. (\ref{beta0}) in Eq. (\ref{eq:ueff}) we get  
\begin{equation}
 u_{\rm eff}(\beta_0)= {k_0 \over 2} - {k_{-1}^2 \over 8 k_{-2}}.
\end{equation} 
 For $\delta=\lambda=0$ this takes the form 
\begin{equation}\label{depth}
 u_{\rm eff}(\beta_0)= {a^2 \over 2} \left({25\over 4}+\Lambda \right)- {(1-4a -a\Lambda)^2 \over 4+2\Lambda + 4 \tilde B}.
\end{equation}

\subsection*{7.4 Comparison to the usual Kratzer potential}

The usual Kratzer potential (without deformation dependence of the mass) in rescaled notation is given in Eq. (\ref{Br}).
Equating the derivative of the effective potential to zero we easily find 
the position, $\tilde \beta_0$, of the minimum of the potential
\begin{equation}
\tilde \beta_0= {2\tilde B_r\over A},  
\end{equation}
which in non-rescaled notation takes the form
\begin{equation}
\beta_0= 2\tilde B,  
\end{equation}
where use of Eqs. (\ref{tildeb}) and  (\ref{Br}) has been made. 

We see that this expression is in agreement with Eq. (\ref{beta0})
in the case of $a=0$ and $\tilde B >> 1$. 

As mentioned above, as depth of the potential its value at the position of its minimum is considered. 
We then easily find 
\begin{equation}
u(\tilde \beta_0)= -{A^2 \over 4\tilde B_r},
\end{equation}
which in non-rescaled notation takes the form
\begin{equation}
u(\beta_0)= -{1 \over 4\tilde B},
\end{equation}
where use of Eqs. (\ref{tildeb}) and (\ref{Br}) has been made. 

We see that this expression is in agreement with Eq. (\ref{depth})
in the case of $a=0$ and $\tilde B >> 1$. 

\subsection*{7.5 Wave functions}

The wave functions $\xi(\beta)$ are normalized through 
\begin{equation}
\int_0^\infty \beta^4 |\xi(\beta)|^2 d\beta=1, 
\end{equation}
while the rescaled wave functions $\xi_r(\tilde \beta)$ are normalized through 
\begin{equation}
\int_0^\infty \tilde \beta^4 |\xi_r(\tilde \beta)|^2 d\tilde \beta=1. 
\end{equation}
From these relations one gets
\begin{equation}
\xi_r(\tilde \beta)= {\xi(\beta)\over A^{5/2}}. 
\end{equation}

The wave functions $R(\beta)$ are normalized through 
\begin{equation}
\int_0^\infty  |R(\beta)|^2 d\beta=1, 
\end{equation}
while the rescaled wave functions $R_r(\tilde \beta)$are normalized through 
\begin{equation}
\int_0^\infty |R_r(\tilde \beta)|^2 d\tilde \beta=1. 
\end{equation}
From these relations one gets
\begin{equation}\label{Rresc}
R_r(\tilde \beta)= {R(\beta)\over A^{1/2}}. 
\end{equation}

These results are consistent with Eq. (\ref{xiR}), leading to 
\begin{equation}
\xi_r(\tilde \beta)= {R_r(\tilde \beta)\over \tilde \beta^2}.  
\end{equation}

These results can also be checked against the explicit expressions for the wave functions.

In Eq. (\ref{RR0}) we see that $R_0(\beta)$ will scale with $N_0 \beta^{-\mu}$, 
while from Eq. (\ref{Nzero}) we see that $N_0$ will scale as $a^{-\mu +{1\over 2}}$. 
Therefore we get 
\begin{equation}
R_0(\beta)\propto a^{-\mu +{1\over 2}} \beta^{-\mu}
\end{equation}
and then in the rescaled framework 
\begin{eqnarray}
R_{0,r}(\tilde \beta)\propto a_r^{-\mu +{1\over 2}} \tilde \beta^{-\mu} = \left({a\over A}\right)^{-\mu +{1\over 2}} \tilde (A \beta)^{-\mu}
\nonumber \\
\propto A^{-1/2} R_0(\beta), 
\end{eqnarray}
in agreement with  Eq. (\ref{Rresc}). 

Similarly, in Eq. (\ref{RRn}) we see that $R_n(\beta)$ will scale with $N_n \beta^{-\mu_n}$, 
while from Eq. (\ref{Nall}) we see that $N_n$ will scale as $a^{-\mu_n +{1\over 2}}$. 
Therefore we get 
\begin{equation}
R_n(\beta)\propto a^{-\mu_n +{1\over 2}} \beta^{-\mu_n}
\end{equation}
and then in the rescaled framework 
\begin{eqnarray}
R_{n,r}(\tilde \beta)\propto a_r^{-\mu_n +{1\over 2}} \tilde \beta^{-\mu_n} = \left({a\over A}\right)^{-\mu_n +{1\over 2}} (A \beta)^{-\mu_n}
\nonumber \\
\propto A^{-1/2} R_n(\beta), 
\end{eqnarray}
in agreement with  Eq. (\ref{Rresc}). 

\subsection*{7.6 $B(E2)$s}

$B(E2)$s are given by Eq. (B4) of Ref. \cite{ESDPRC}, in which the square of the radial integral appears.
The radial integral for $\gamma$-unstable nuclei is given in Eq. (116) of \cite{DDMD}
\begin{equation}
\begin{split}
& I_{n',\tau+1; n,\tau} = \int_0^\infty \beta \xi_{n',\tau+1}(\beta) \xi_{n,\tau}(\beta) \beta^4 d\beta \\
& =\int_0^\infty \beta R_{n',\tau+1}(\beta) R_{n,\tau}(\beta) d\beta.
\end{split}
\end{equation}
From Appendix 7.5 it is clear that $I$ behaves in scaling as $\beta$. In other words, in the rescaled framework 
we are going to have 
\begin{equation}
I_r = A I.
\end{equation}
Therefore in the rescaled framework for the $B(E2)$s we are going to have 
\begin{equation}\label{BE2scale}
B(E2)_r = A^2 B(E2). 
\end{equation}
For prolate deformed nuclei, the radial integral is given by Eq. (117) of Ref. \cite{DDMD}.
The same scaling properties occur then.  

\subsection*{7.7 Numerical results}

The numerical results on the spectra, reported in Tables I-IV, are not affected by rescaling since, according to Eq. (\ref{Escale}),
energies scale as $A^2$, which cancels out when energy ratios are calculated. 

Furthermore the numerical results on the $B(E2)$s, reported in Tables V and VI, are not affected by rescaling, since 
the powers of $A$ appearing because of the rescaling of wave functions, as described in Appendix 7.5,
cancel out when $B(E2)$ ratios are calculated, as it is clear from Appendix 7.6. 

The only results affected by scaling are the figures of the effective potential.
From Eqs. (\ref{eff1}) and (\ref{eff2}) it is clear that the numerical values of the effective potentials 
are the same, irrespectively from calculating them in the rescaled framework or in the non-rescaled framework.
Then the only rescaling entering is by an overall factor of $A^2$, as seen in Eq. (\ref{Vscale}). 

What is affected, however, is the abscissa. If the non-rescaled quantity $\beta$ is used, it obtains very high numerical values, 
outside the region of physical interest. 
Here the rescaling procedure comes in. One way to obtain the value of the rescaling factor $A$ for each nucleus 
is to use it in order to have the minimum of the effective potential at the value of $\beta$ corresponding 
to the quadrupole deformation of the specific nucleus obtained from $B(E2;0_1^+\to 2_1^+)$, i.e. from the transition rate 
from the ground state to the first excited state \cite{Raman}. This can be done using Eq. (\ref{Aunst}) for $\gamma$-unstable nuclei, 
or Eq. (\ref{Aprol}) for prolate deformed nuclei. 

Numerical values of the rescaling parameter $A$ for the $\gamma$-unstable nuclei shown in Figs. 2 and 3  
are shown in Table VII,  while numerical values of $A$ for the prolate deformed nuclei shown in Figs. 4, 5 and 6  
are shown in Table VIII. Figs 2-6 have been drawn using these values of $A$. The potential values have been divided by $A^2$, 
as indicated by Eq. (\ref{Vscale}), while in the abscissa $\beta$ has been multiplied by $A$, according to Eq. (\ref{tildeb}).   

As an extra qualitative check, the depth of each potential, which  
should have been approximately equal to $-1/(4\tilde B)$ in a non-rescaled version of the figures 2-6, 
as discussed in Appendices 7.3 and 7.4, should be $-A^2/(4\tilde B)$ in the rescaled version. 
For better accuracy, Eq. (\ref{depth}) should be used. 
   
It should be noticed that the scale appropriate for each individual nucleus can be fixed in several ways.

1) Since energy levels scale as $A^2$, as seen in Eq. (\ref{Escale}), one way to determine $A$ is by fitting it to
the energy of the first excited state, $E(2_1^+)$, which is readily available for many nuclei \cite{ENSDF}. 

2) Since $B(E2)$s also scale as $A^2$, as seen in Eq. (\ref{BE2scale}), another way to determine $A$ is by fitting it to
the transition rate from the ground state to the first excited state, $B(E2; 0_1^+\to 2_1^+)$, which is also 
readily available for many nuclei \cite{Raman}. 

The method used here, namely fitting the minimum of the effective potential to the quadrupole deformation 
found from $B(E2; 0_1^+\to 2_1^+)$, has the advantage that it avoids units, since $\beta$ is a dimensionless quantity. 
It also avoids the scale factor appearing in the expression for the transition operator (see Eq. (115) of Ref. \cite{DDMD}).

\newpage

\begin{table*}

\caption{Comparison of theoretical predictions of the 
$\gamma$-unstable Bohr Hamiltonian with $\beta$-dependent mass 
(with $\delta =\lambda =0$) for the Kratzer potential 
to experimental data \cite{NDS} of
rare earth and actinides with $R_{4/2} \leq 2.6$ and known $0_2^+$ 
and $2_{\gamma}^+$ states.  The $R_{4/2}=E(4_1^+)/E(2_1^+)$
ratios, as well as the $\beta$ and $\gamma$ bandheads, normalized
to the $2_1^+$ state and labelled by
$R_{0/2}=E(0_{\beta}^+)/E(2_1^+)$ and
$R_{2/2}=E(2_{\gamma}^+)/E(2_1^+)$ respectively, are shown. The
angular momenta of the highest levels of the ground state, $\beta$
and $\gamma$ bands included in the rms fit are labelled by $L_g$,
$L_\beta$, and $L_\gamma$ respectively, while $n$ indicates the
total number of levels involved in the fit and $\sigma$ is the
quality measure of Eq. (\ref{eq:e99}). The theoretical predictions are obtained 
from the formulae mentioned in Sec. \ref{enespe}.
See subsec. \ref{specgam} for further discussion. }

\bigskip

\begin{tabular}{ r r r r  r r r r  r r r r r r}
\hline 
 & & & & & & & & & & & & & \\
nucleus & $R_{4/2}$ & $R_{4/2}$ & $R_{0/2}$& $R_{0/2}$
&$R_{2/2}$ & $R_{2/2}$ & $\tilde B$ & $a$ &
$L_g$ & $L_\beta$ & $L_\gamma$ & $n$ & $\sigma$ \\
        & exp &  th  & exp & th  & exp & th &  &  &  &  &  &   &  \\

\hline

$^{98}$Ru  & 2.14 & 2.34 & 2.0 &  2.0 &  2.2 &  2.3 &  38 & 0.0101 
& 24 &   0 &   4 & 15 & 0.811 \\
$^{100}$Ru & 2.27 & 2.40 & 2.1 &  2.1 &  2.5 &  2.4 &  64 & 0.0094
& 24 &   0 &   4 & 15 & 0.824 \\

$^{102}$Ru & 2.33 & 2.35 & 2.0 &  2.0 &  2.3 &  2.3 &  41 & 0.0105
& 16 &   0 &   5 & 12 & 0.384 \\
$^{104}$Ru & 2.48 & 2.39 & 2.8 &  3.1 &  2.5 &  2.4 &  60 & 0.0046
&  8 &  2  &   8 & 12 & 0.437 \\ 

$^{102}$Pd & 2.29 & 2.42 & 2.9 & 2.4 & 2.8 & 2.4 &  81 & 0.0072
& 24 &  4  &  4 & 17 & 0.996 \\
$^{104}$Pd & 2.38 & 2.35 & 2.4 & 2.4 & 2.4 & 2.3 &  41 & 0.0072
& 18 &  2  &  4 & 13 & 0.328 \\
$^{106}$Pd & 2.40 & 2.33 & 2.2 & 2.3 & 2.2 & 2.3 &  36 & 0.0082
& 16 &  4  &  5 & 14 & 0.398 \\
$^{108}$Pd & 2.42 & 2.38 & 2.4 & 2.5 & 2.1 & 2.4 &  55 & 0.0069
& 14 &  4  &  4 & 12 & 0.317 \\
$^{110}$Pd & 2.46 & 2.43 & 2.5 & 2.7 & 2.2 & 2.4 & 100 & 0.0061 
& 12 &  10 &  4 & 14 & 0.377 \\
$^{112}$Pd & 2.53 & 2.32 & 2.6 & 2.6 & 2.1 & 2.3 &  33 & 0.0058 
&  6 &   0 &  3 &  5 & 0.485 \\
$^{114}$Pd & 2.56 & 2.40 & 2.6 & 2.6 & 2.1 & 2.4 &  65 & 0.0065 
& 16 &   0 & 11 & 18 & 0.772 \\
$^{116}$Pd & 2.58 & 2.42 & 3.3 & 3.3 & 2.2 & 2.4 &  83 & 0.0044 
& 16 &   0 &  9 & 16 & 0.630 \\

$^{106}$Cd & 2.36 & 2.33 & 2.8 & 2.8 & 2.7 & 2.3 &  36 & 0.0044 
& 12 & 0 & 2 &  7 & 0.174 \\
$^{108}$Cd & 2.38 & 2.34 & 2.7 & 2.7 & 2.5 & 2.3 &  39 & 0.0054 
& 22 & 0 & 5 & 15 & 0.908 \\
$^{110}$Cd & 2.35 & 2.29 & 2.2 & 1.9 & 2.2 & 2.3 &  28 & 0.0115 
& 16 & 6 & 5 & 15 & 0.341 \\
$^{112}$Cd & 2.29 & 2.23 & 2.0 & 1.7 & 2.1 & 2.2 &  20 & 0.0126 
& 12 & 8 &11 & 20 & 0.282 \\
$^{114}$Cd & 2.30 & 2.25 & 2.0 & 1.7 & 2.2 & 2.2 &  22 & 0.0127
& 14 & 4 & 3 & 11 & 0.249 \\
$^{116}$Cd & 2.38 & 2.27 & 2.5 & 2.8 & 2.4 & 2.3 &  25 & 0.0028
& 14 & 2 & 3 & 10 & 0.306 \\ 
$^{118}$Cd & 2.39 & 2.29 & 2.6 & 2.6 & 2.6 & 2.3 &  28 & 0.0045
& 14 & 0 & 3 &  9 & 0.312 \\
$^{120}$Cd & 2.38 & 2.31 & 2.7 & 2.7 & 2.6 & 2.3 &  32 & 0.0045
& 16 & 0 & 2 &  9 & 0.426 \\

$^{118}$Xe & 2.40 & 2.41 & 2.5 & 2.8 & 2.8 & 2.4 &  77 & 0.0058
& 16 & 4 &10 & 19 & 0.408 \\
$^{120}$Xe & 2.47 & 2.45 & 2.8 & 3.0 & 2.7 & 2.4 & 133 & 0.0049
& 26 & 4 & 9 & 23 & 0.701 \\
$^{122}$Xe & 2.50 & 2.45 & 3.5 & 3.5 & 2.5 & 2.4 & 131 & 0.0040 
& 16 & 0 & 9 & 16 & 0.731 \\
$^{124}$Xe & 2.48 & 2.43 & 3.6 & 3.7 & 2.4 & 2.4 &  93 & 0.0035 
& 20 & 2 &11 & 21 & 0.722 \\
$^{126}$Xe & 2.42 & 2.39 & 3.4 & 3.4 & 2.3 & 2.4 &  60 & 0.0035 
& 12 & 4 & 9 & 16 & 0.601 \\
$^{128}$Xe & 2.33 & 2.31 & 3.6 & 3.7 & 2.2 & 2.3 &  32 & 0.0000 
& 10 & 2 & 7 & 12 & 0.451 \\
$^{130}$Xe & 2.25 & 2.30 & 3.3 & 3.3 & 2.1 & 2.3 &  29 & 0.0007 
& 14 & 0 & 5 & 11 & 0.477 \\
$^{132}$Xe & 2.16 & 2.03 & 2.8 & 2.0 & 1.9 & 2.0 &   9 & 0.0000 
&  6 & 0 & 5 &  7 & 0.374 \\
$^{134}$Xe & 2.04 & 1.87 & 1.9 & 1.6 & 1.9 & 1.9 &   5 & 0.0000 
&  6 & 0 & 5 &  7 & 0.216 \\

$^{130}$Ba & 2.52 & 2.45 & 3.3 & 3.3 & 2.5 & 2.4 & 140 & 0.0043 
& 12 & 0 & 6 & 11 & 0.392 \\  
$^{132}$Ba & 2.43 & 2.37 & 3.2 & 3.2 & 2.2 & 2.4 &  50 & 0.0037 
& 14 & 0 & 8 & 14 & 0.763 \\
$^{134}$Ba & 2.32 & 2.20 & 2.9 & 2.8 & 1.9 & 2.2 &  18 & 0.0000 
&  8 & 0 & 4 &  7 & 0.344 \\
$^{136}$Ba & 2.28 & 2.00 & 1.9 & 1.9 & 1.9 & 2.0 &   8 & 0.0002 
&  6 & 0 & 2 &  4 & 0.192 \\  
$^{142}$Ba & 2.32 & 2.41 & 4.3 & 4.3 & 4.0 & 2.4 &  79 & 0.0021 
& 14 & 0 & 2 &  8 & 0.591 \\

$^{134}$Ce & 2.56 & 2.42 &  3.7 &  3.9 &  2.4 & 2.4 &   88 & 0.0030 &
28 & 2 & 8 & 22 & 0.882 \\
$^{136}$Ce & 2.38 & 2.28 &  1.9 &  1.9 &  2.0 & 2.3 &   27 & 0.0105 &
16 & 0 & 3 & 10 & 0.546 \\
$^{138}$Ce & 2.32 & 2.13 &  1.9 &  1.9 &  1.9 & 2.1 &   13 & 0.0083 &
14 & 0 & 2 &  8 & 0.350 \\

$^{140}$Nd & 2.33 & 2.09 &  1.8 &  1.8 & 1.9 & 2.1 &  11 & 0.0073 &
 6 & 0 & 2 &  4 & 0.168 \\
$^{148}$Nd & 2.49 & 2.42 &  3.0 &  3.3 & 4.1 & 2.4 &  90 & 0.0042 &
12 & 8 & 4 & 13 & 0.719 \\

$^{140}$Sm & 2.35 & 2.36 & 1.9 & 1.9 & 2.7 & 2.4 &  44 & 0.0115 &
 8 & 0 & 2 &  5 & 0.161 \\
$^{142}$Sm & 2.33 & 2.16 & 1.9 & 1.9 & 2.2 & 2.2 &  15 & 0.0089 &
 8 & 0 & 2 &  5 & 0.114 \\

$^{142}$Gd & 2.35 & 2.33 & 2.7 & 2.6 & 1.9 & 2.3 &  35 & 0.0054 &
16 & 0 & 2 &  9 & 0.290 \\
$^{144}$Gd & 2.35 & 2.35 & 2.5 & 2.5 & 2.5 & 2.3 &  41 & 0.0065 &
 6 & 0 & 2 &  4 & 0.108 \\
$^{152}$Gd & 2.19 & 2.34 & 1.8 & 1.9 & 3.2 & 2.3 &  40 & 0.0116 &
16 & 10& 7 & 19 & 0.382 \\
 
$^{154}$Dy  & 2.23 & 2.40 & 2.0 & 1.7 & 3.1 & 2.4 &  67 & 0.0124 &
26 & 10 & 7 & 24 & 0.948 \\

$^{156}$Er  & 2.32 & 2.38 & 2.7 & 2.7 & 2.7 & 2.4 &  56 & 0.0062 &
20 & 4 & 5 & 16 & 0.357 \\

\hline
\end{tabular}
\end{table*}

\begin{table*}
\setcounter{table}{0} \caption{(continued)}

\bigskip

\begin{tabular}{ r r r r  r r r r  r r r r r r}
\hline 
 & & & & & & & & & & & & & \\
nucleus & $R_{4/2}$ & $R_{4/2}$ & $R_{0/2}$& $R_{0/2}$
&$R_{2/2}$ & $R_{2/2}$ & $\tilde B$ & $a$ & 
$L_g$ & $L_\beta$ & $L_\gamma$ & $n$ & $\sigma$ \\
        & exp &  th  & exp & th  & exp & th &  &  &  &  &    & &  \\

\hline

$^{186}$Pt & 2.56 & 2.47 & 2.5 & 3.6 & 3.2 & 2.5 & 249 & 0.0035 &
26 & 6 & 10 & 25 & 0.791 \\ 
$^{188}$Pt & 2.53 & 2.43 & 3.0 & 3.2 & 2.3 & 2.4 & 100 & 0.0047 &
16 & 2 & 4 & 12 & 0.455 \\
$^{190}$Pt & 2.49 & 2.37 & 3.1 & 3.2 & 2.0 & 2.4 &  49 & 0.0038 &
18 & 2 & 6 & 15 & 0.538 \\
$^{192}$Pt & 2.48 & 2.38 & 3.8 & 3.8 & 1.9 & 2.4 &  53 & 0.0021 &
10 & 0 & 8 & 12 & 0.698 \\
$^{194}$Pt & 2.47 & 2.39 & 3.9 & 3.9 & 1.9 & 2.4 &  60 & 0.0023 &
10& 4 & 5 & 11 & 0.688 \\
$^{196}$Pt & 2.47 & 2.38 & 3.2 & 3.1 & 1.9 & 2.4 &  54 & 0.0043 &
10 & 2 & 6 & 11 & 0.676 \\ 
$^{198}$Pt & 2.42 & 2.25 & 2.2 & 2.3 & 1.9 & 2.3 &  23 & 0.0059 &
 6 & 2 & 4 & 7  & 0.372 \\
$^{200}$Pt & 2.35 & 2.00 & 2.4 & 1.9 & 1.8 & 2.0 &   8 & 0.0000 &
 4 & 0 & 4 &  5 & 0.342 \\

\hline
\end{tabular}
\end{table*}

\newpage

\begin{table*}

\caption{Comparison of theoretical predictions of the Bohr Hamiltonian with
$\beta$-dependent mass (with $\delta =\lambda =0$) for the Kratzer potential 
for axially symmetric prolate deformed nuclei 
to experimental data \cite{NDS} of
rare earth and actinides with $R_{4/2}$ $>$ 2.9 and known $0_2^+$
and $2_{\gamma}^+$ states.  The $R_{4/2}=E(4_1^+)/E(2_1^+)$
ratios, as well as the $\beta$ and $\gamma$ bandheads, normalized
to the $2_1^+$ state and labelled by
$R_{0/2}=E(0_{\beta}^+)/E(2_1^+)$ and
$R_{2/2}=E(2_{\gamma}^+)/E(2_1^+)$ respectively, are shown. The
angular momenta of the highest levels of the ground state, $\beta$
and $\gamma$ bands included in the rms fit are labelled by $L_g$,
$L_\beta$, and $L_\gamma$ respectively, while $n$ indicates the
total number of levels involved in the fit and $\sigma$ is the
quality measure of Eq. (\ref{eq:e99}). 
The theoretical predictions are obtained from the equations mentioned in Sec. \ref{enespe}.
See subsec. \ref{specdef} for further
discussion. }

\bigskip

\begin{tabular}{ l r r r  r r r r r r r r r r r}
\hline 
 & & & & & & & & & & & & & & \\
nucleus & $R_{4/2}$ & $R_{4/2}$ & $R_{0/2}$& $R_{0/2}$
&$R_{2/2}$ & $R_{2/2}$ & $\tilde B$ & $c$   & $a$ &
$L_g$ & $L_\beta$ & $L_\gamma$ & $n$ & $\sigma$ \\
        & exp &  th  & exp & th  & exp & th &  &  &  & & &  &   &  \\

\hline

$^{150}$Nd & 2.93 & 3.17 & 5.2 &  6.1 &  8.2 &  8.5 &   31 & 3.3 &  0.0033 
& 14 &   6 &   4 & 13 & 0.655 \\
 
$^{152}$Sm & 3.01 & 3.22 & 5.6 &  5.5 &  8.9 &  10.0 &  48 & 3.8 & 0.0050
& 16 &  14 &   9 & 23 & 0.622 \\
$^{154}$Sm & 3.25 & 3.29 & 13.4 & 13.6 & 17.6 & 18.6 & 144 & 6.8 & 0.0007
& 16 &   6 &   7 & 17 & 0.503 \\

$^{154}$Gd & 3.02 & 3.23 &  5.5 &  5.3 &  8.1 &  8.4 &  62  & 3.0 & 0.0056
& 20 & 20  &   7 & 26 & 0.926 \\ 
$^{156}$Gd & 3.24 & 3.29 & 11.8 & 11.3 & 13.0 & 13.7 & 159  & 4.8 & 0.0014
& 26 & 12  &   16 & 34 & 0.973 \\
$^{158}$Gd & 3.29 & 3.30 & 15.0 & 14.8 & 14.9 & 15.2 & 202 & 5.3 & 0.0007
& 12 &  6  &   6 & 14 & 0.149 \\
$^{160}$Gd & 3.30 & 3.31 & 17.6 & 17.6 & 13.1 & 13.0 & 287 & 4.4 & 0.0005
& 16 &  4  &   8 & 17 & 0.141 \\
$^{162}$Gd & 3.29 & 3.31 & 19.8 & 19.9 & 12.0 & 11.9 & 261 & 4.0 & 0.0002
& 14 &  0  &   4 & 10 & 0.097 \\

$^{156}$Dy & 2.93 & 3.21 &  4.9 &  5.1 &  6.5 &  6.6 &  51  & 2.3 & 0.0060 & 
20 &  10 &  13 & 27 & 0.832 \\
$^{158}$Dy & 3.21 & 3.27 & 10.0 &  9.9 &  9.6 & 10.1 & 113  & 3.5 & 0.0017 & 
24 &  8 &  8 & 23 & 0.830 \\
$^{160}$Dy & 3.27 & 3.29 & 14.7 & 14.7 & 11.1 & 11.4 & 176 & 3.9 & 0.0006 & 
28 & 4 & 23 & 38 & 0.927 \\
$^{162}$Dy & 3.29 & 3.30 & 17.3 & 15.5 & 11.0 & 11.0 & 247 & 3.7 & 0.0007 &
18 &  10 & 14 & 27 & 0.830 \\
$^{164}$Dy & 3.30 & 3.31 & 22.6 & 22.9 & 10.4 & 10.2 & 281 & 3.4 & 0.0000 & 
20 & 0 & 10 & 19 & 0.199 \\
$^{166}$Dy & 3.31 & 3.30 & 15.0 & 15.0 & 11.2 & 11.2 & 214 & 3.8 & 0.0007 & 
6 & 2 & 5 & 8 & 0.060 \\

$^{160}$Er & 3.10 & 3.24 &  7.1 &  7.2 &  6.8 & 6.9 &  65 &  2.4 & 0.0031
& 22 & 2 & 5 & 16 & 0.874\\
$^{162}$Er & 3.23 & 3.27 & 10.7  & 10.6 &  8.8 & 9.7 & 100 & 3.4 & 0.0012
&20 & 4 & 12 & 23 & 0.518 \\ 
$^{164}$Er & 3.28 & 3.29 & 13.6 & 12.9 &  9.4 &  9.2 & 179 & 3.1 & 0.0010
&22 & 10 & 19 & 34 & 0.915 \\
$^{166}$Er & 3.29 & 3.29 & 18.1 & 17.6 &  9.8 & 10.0 & 167 &  3.4 & 0.0000
& 16 & 10 & 14 & 26 & 0.340 \\
$^{168}$Er & 3.31 & 3.31 & 15.3 & 14.5 & 10.3 & 10.3 & 384 & 3.4 & 0.0010
& 18 & 6 & 8 & 19 & 0.274 \\
$^{170}$Er & 3.31 & 3.32 & 11.3 &  9.9 & 11.9 & 12.4 & 491 & 4.1 & 0.0018
&26 & 16 & 19 & 39 & 0.807 \\

$^{162}$Yb & 2.92 & 3.18 & 3.6 & 3.6 &   4.8 & 5.0 &  40 &  1.7 & 0.0103 &
20 & 0 & 4 & 13 & 0.944\\
$^{164}$Yb & 3.13 & 3.24 & 7.9 & 7.9 &   7.0 & 7.2 &  72 &  2.5 & 0.0025 &
18 & 0 & 5 & 13 & 0.771\\
$^{166}$Yb & 3.23& 3.28 & 10.2 & 9.6 &    9.1 & 8.9 & 138 &  3.0 & 0.0020 &
24 & 10 & 13 & 29 & 0.974 \\
$^{168}$Yb & 3.27 & 3.29 & 13.2& 13.0 & 11.2 &11.3 & 160 & 3.9 & 0.0009 &
24 & 4 & 7 & 20 & 0.710 \\
$^{170}$Yb & 3.29 & 3.29 & 12.7 & 11.1 & 13.6 &14.0 & 172 & 4.9 & 0.0015 &
20 & 18 & 17 & 35 & 0.822 \\
$^{172}$Yb & 3.31 & 3.31 & 13.2 & 12.7 & 18.6 & 18.8 & 246  & 6.6 & 0.0012 &
16 & 14 & 5 & 19 & 0.787 \\
$^{174}$Yb & 3.31 & 3.32 & 19.4 & 19.1 & 21.4 & 21.5 & 398 & 7.4 & 0.0005 &
20 & 4 & 5 & 16 & 0.208 \\
$^{176}$Yb & 3.31 & 3.31 & 13.9 &  13.5 & 15.4 & 15.5 & 296 & 5.3 & 0.0011 &
20 & 2 & 5 & 15 & 0.129 \\  
$^{178}$Yb & 3.31 & 3.31 & 15.7 &  15.6 & 14.5 & 14.6 & 254 & 5.0 & 0.0007 & 
6 & 4 & 2 & 6 & 0.025 \\

$^{166}$Hf & 2.97 & 3.19 &  4.4 &  4.4 &  5.1 & 5.3 &   44 & 1.8 & 0.0079 &
20 & 0 & 3 & 12 & 0.983\\
$^{168}$Hf & 3.11 & 3.25 &  7.6 &  7.6 &  7.1 & 7.6 &   80 & 2.6 & 0.0029 &
22 & 4 & 4 & 16 & 1.043\\
$^{170}$Hf & 3.19 & 3.27 &  8.7 &  8.8 &  9.5 & 10.0 &  99 & 3.5 & 0.0022 &
22 & 4 & 4 & 16 & 0.928 \\
$^{172}$Hf & 3.25 & 3.29 &  9.2 &  9.0 & 11.3 & 11.6 & 150  & 4.0 & 0.0023 &
26 & 4 & 6 & 20 & 0.996 \\
$^{174}$Hf & 3.27 & 3.29 &  9.1 &  7.7 & 13.5 & 13.9 & 154 & 4.9 & 0.0030 &
24 & 20 & 5 & 26 & 1.005 \\
$^{176}$Hf & 3.28 & 3.30 & 13.0 & 12.3 & 15.2 & 15.9 & 190 & 5.6 & 0.0012 &
18 & 10 & 8 & 21 & 0.569 \\
$^{178}$Hf & 3.29 & 3.29 & 12.9 & 12.9 & 12.6 & 13.0 & 172 & 4.5 & 0.0010 &
18 & 6 & 6 & 17 & 0.141 \\
$^{180}$Hf & 3.31 & 3.31 & 11.8 & 11.6 & 12.9 & 12.8 & 350 & 4.3 & 0.0015 &
12 & 4 & 5 & 12 & 0.121 \\

$^{176}$W  & 3.22 & 3.27 & 7.8 &  7.3 &  9.6 & 10.3 & 104 &  3.6 & 0.0033 &
22& 12 & 5 & 21 & 0.811\\
$^{178}$W  & 3.24 & 3.27&  9.4 &  8.9 & 10.5 & 10.5 &  97 &  3.7 & 0.0021 &
 18 & 14 & 2 & 17 & 0.356 \\
$^{180}$W  & 3.26 & 3.28 &  14.6 & 14.6 & 10.8 & 11.4 & 118 &  4.0 & 0.0001 &
24 & 0 & 7 & 18 & 0.832 \\
$^{182}$W  & 3.29 & 3.31 & 11.3& 11.5 & 12.2 & 12.4 & 256 &  4.2 & 0.0015 &
18 & 4 & 6 & 16 & 0.189 \\
$^{184}$W  & 3.27 & 3.29 & 9.0 &  9.1 &  8.1 &  8.1 & 164 &  2.7 & 0.0023 &
 10 & 4 & 6 & 12 & 0.091 \\ 
$^{186}$W  & 3.23 & 3.29 &  7.2 &  7.5 &  6.0 &  6.1 & 148 &  2.0 & 0.0033 &
14 & 4 & 6 & 14 & 0.156 \\

$^{176}$Os & 2.93 & 3.19 &  4.5 & 4.9 & 6.4 & 7.0 &  42 & 2.5 & 0.0063 &
18& 6 & 5 & 16 & 0.984 \\
$^{178}$Os & 3.02 & 3.20 &  4.9 & 5.2 & 6.6 & 7.2 & 42 & 2.6 & 0.0056 &
16& 6 & 5 & 15 & 0.636 \\
$^{180}$Os & 3.09 & 3.20 &  5.6 & 6.7 & 6.6 & 7.4 &  43 & 2.7 & 0.0030 &
14& 4 & 7 & 15 & 0.911 \\
$^{184}$Os & 3.20 & 3.26 &  8.7 & 8.7 &  7.9 & 8.4 & 91 & 2.9 & 0.0022 &
22 & 0 & 6 & 16 & 0.452 \\ 
$^{186}$Os & 3.17 & 3.25 &  7.7 & 7.7 &  5.6 & 6.0 & 84 & 2.0 & 0.0029 &
14 & 10 & 13 & 24 & 0.249 \\
$^{188}$Os & 3.08 & 3.21 &  7.0 & 7.3 &  4.1 & 4.3 & 50 & 1.4 & 0.0023 &
12 & 2 & 7 & 13 & 0.214 \\
$^{190}$Os & 2.93 & 3.13 &  4.9 & 5.0 &  3.0 & 3.1 & 30 & 1.0 & 0.0054 &
10 & 2 & 6 & 11 & 0.230 \\

\hline
\end{tabular}
\end{table*}

\begin{table*}
\setcounter{table}{1} \caption{(continued)}

\bigskip

\begin{tabular}{ l r r r  r r r r r r r r r r r}
\hline 
 & & & & & & & & & & & & & & \\
nucleus & $R_{4/2}$ & $R_{4/2}$ & $R_{0/2}$& $R_{0/2}$
&$R_{2/2}$ & $R_{2/2}$ & $\tilde B$ & $c$   & $a$ & 
$L_g$ & $L_\beta$ & $L_\gamma$ & $n$ & $\sigma$ \\
        & exp &  th  & exp & th  & exp & th &  &  &  &  &  &  & &  \\

\hline

$^{228}$Ra & 3.21 & 3.28 & 11.3 & 11.1 & 13.3 & 13.4 & 116 & 4.8 & 0.0012 
& 22 & 4 & 3 & 15 & 0.706 \\

$^{228}$Th & 3.24 & 3.28 & 14.4 & 14.2 & 16.8 & 17.1 & 120 & 6.3 & 0.0003 
& 18 & 2 & 5 & 14 & 0.396 \\
$^{230}$Th & 3.27 & 3.30 & 11.9 & 11.7 & 14.7 &  14.7 & 213  & 5.1 & 0.0014
& 24 & 4 & 4 & 17 & 0.625 \\
$^{232}$Th & 3.28 & 3.31 & 14.8 & 14.5 & 15.9 & 16.0 & 268 & 5.5 & 0.0009
& 30 & 20 & 12 & 36 & 0.964 \\
 
$^{232}$U  & 3.29 & 3.30 & 14.5 & 14.8 & 18.2 & 18.2 & 234 & 6.4 & 0.0008
&20 & 10 & 4 & 18 & 0.244 \\
$^{234}$U  & 3.30 & 3.31 & 18.6 & 19.1 & 21.3 & 21.4 & 307 & 7.5 & 0.0004
& 28 & 8 & 7 & 24 & 0.785 \\
$^{236}$U  & 3.30 & 3.31 & 20.3& 19.8 & 21.2 & 21.3 & 354 & 7.4 & 0.0004
& 30 & 4 & 5 & 21 & 0.700 \\
$^{238}$U  & 3.30 & 3.31 & 20.6 & 20.2 & 23.6 & 24.4 & 378 & 8.5 & 0.0004
& 30 & 4 & 27 & 43 & 0.911 \\

$^{238}$Pu & 3.31 & 3.32 & 21.4 & 21.5 & 23.3 & 23.3 & 498 & 8.0 &  0.0004
&26 & 2 & 4 & 17 & 0.368 \\
$^{240}$Pu & 3.31 & 3.32 & 20.1 & 19.6 & 26.6 & 26.7 & 452 & 9.3 & 0.0005
& 26 & 4 & 4 & 18 & 0.516 \\
$^{242}$Pu & 3.31 & 3.32 & 21.5 & 20.8 & 24.7 & 24.8 & 422 & 8.6 & 0.0004
& 26 & 2 & 2 & 15 & 0.402 \\

$^{248}$Cm & 3.31 & 3.32 & 25.0 & 24.3 & 24.2 & 24.2 & 429 & 8.4 & 0.0002
&28 & 4 & 2 & 17 & 0.458 \\

$^{250}$Cf & 3.32 & 3.31& 27.0 & 27.0 & 24.2 & 24.1 & 375 & 8.4 & 0.0000
&8 & 2 & 4 & 8 & 0.078 \\

\hline
\end{tabular}
\end{table*}

\begin{table*}

\caption{Normalized [to the energy of the first excited state, $E(2_1^+)$] energy levels of the ground state band (gsb) 
and the  $\beta_1$ and $\gamma_1$ bands of $^{170}$Er and $^{232}$Th,
obtained from the Bohr Hamiltonian with $\beta$-dependent mass for the Kratzer potential for axially symmetric prolate deformed nuclei
using the parameters given in Table~II, compared to experimental data \cite{NDS}. See subsec. \ref{specdef}
for further discussion. }

\bigskip

\begin{tabular}{  r r r r r r r r r r r r r r}
\hline 
 & & & & & & & & & & & & &  \\
  & $^{170}$Er & $^{170}$Er & $^{232}$Th & $^{232}$Th & $^{170}$Er & $^{170}$Er & $^{232}$Th & $^{232}$Th &  & $^{170}$Er & $^{170}$Er & $^{232}$Th & $^{232}$Th \\ 
L &  exp   & th    & exp   & th    &  exp  & th    & exp   & th    &  L &  exp &   th & exp  & th   \\      

\hline
   &  gsb  &  gsb  &  gsb  &  gsb  & $\beta_1$ & $\beta_1$ & $\beta_1$ & $\beta_1$ & &$\gamma_1$ & $\gamma_1$ & $\gamma_1$ & $\gamma_1$ \\
 0 &  0.00 &  0.00 &  0.00 &  0.00 &  11.3 &   9.9 &  14.8 &  14.5 &  2 & 11.9 & 12.4 & 15.9 & 16.0 \\
 2 &  1.00 &  1.00 &  1.00 &  1.00 &  12.2 &  10.8 &  15.7 &  15.4 &  3 & 12.9 & 13.3 & 16.8 & 16.9 \\
 4 &  3.31 &  3.32 &  3.28 &  3.31 &  14.0 &  13.0 &  17.7 &  17.5 &  4 & 14.3 & 14.6 & 18.0 & 18.0 \\
 6 &  6.88 &  6.92 &  6.75 &  6.86 &  17.2 &  16.3 &  20.7 &  20.7 &  5 & 15.7 & 16.1 & 19.5 & 19.5 \\
 8 & 11.64 & 11.75 & 11.28 & 11.56 &  21.3 &  20.8 &  24.8 &  24.9 &  6 & 17.8 & 18.0 & 21.3 & 21.2 \\
10 & 17.51 & 17.73 & 16.75 & 17.31 &  26.5 &  26.4 &  29.8 &  30.1 &  7 & 19.8 & 20.2 & 23.2 & 23.2 \\
12 & 24.41 & 24.79 & 23.03 & 23.95 &  32.5 &  33.0 &  35.5 &  36.1 &  8 & 22.6 & 22.6 & 25.5 & 25.5 \\
14 & 32.28 & 32.82 & 30.04 & 31.35 &  39.1 &  40.5 &  42.1 &  42.8 &  9 & 25.0 & 25.4 & 27.8 & 28.0 \\
16 & 41.04 & 41.73 & 37.65 & 39.35 &  46.2 &  48.8 &  49.4 &  50.1 & 10 & 28.3 & 28.4 & 30.6 & 30.7 \\
18 & 50.62 & 51.39 & 45.84 & 47.82 &       &       &  57.4 &  57.8 & 11 & 31.1 & 31.6 & 33.2 & 33.6 \\
20 & 60.91 & 61.70 & 55.52 & 56.61 &       &       &  65.8 &  65.8 & 12 & 35.8 & 35.1 & 36.5 & 36.7 \\
22 & 72.20 & 72.55 & 63.69 & 65.60 &       &       &       &       & 13 & 39.1 & 38.9 &      &      \\
24 & 83.80 & 83.82 & 73.32 & 74.67 &       &       &       &       & 14 & 43.7 & 42.8 &      &      \\
26 & 95.82 & 95.42 & 83.38 & 83.73 &       &       &       &       & 15 & 47.2 & 47.0 &      &      \\
28 &       &       & 93.82 & 92.69 &       &       &       &       & 16 & 52.6 & 51.4 &      &      \\
30 &       &       &104.56 &101.49 &       &       &       &       & 17 & 56.2 & 55.9 &      &      \\
   &       &       &       &       &       &       &       &       & 18 & 62.2 & 60.6 &      &      \\
   &       &       &       &       &       &       &       &       & 19 & 66.2 & 65.5 &      &      \\

\hline
\end{tabular}
\end{table*}

\newpage 

\begin{table*}

\caption{Normalized [to the energy of the first excited state, $E(2_1^+)$] energy levels of the ground state band (gsb) 
and the  $\beta_1$ and $\gamma_1$ bands of the ${\rm N}=90$ isotones $^{150}$Nd, $^{152}$Sm, $^{154}$Gd, and $^{156}$Dy,
obtained from the Bohr Hamiltonian with $\beta$-dependent mass for the Kratzer potential for axially symmetric prolate deformed nuclei
using the parameters given in Table~II, compared to experimental data \cite{NDS} and to the predictions of the X(5) critical point symmetry
\cite{IacX5,BonX5,Bijker}. 
The bandhead of the $\gamma_1$ band in X(5), which is a free parameter, has been set equal to the average of the relevant experimental values.
See subsec. \ref{specdef} for further discussion. }

\bigskip

\begin{tabular}{  r r r r r r r r r r}
\hline 
 & & & & & & & & &  \\
  & $^{150}$Nd & $^{150}$Nd & $^{152}$Sm & $^{152}$Sm & $^{154}$Gd & $^{154}$Gd & $^{156}$Dy & $^{156}$Dy & X(5) \\
L &  exp       & th         & exp        &      th    &       exp  &      th    &      exp   &      th    &      \\      

\hline
gsb&       &       &       &       &       &       &       &       &      \\
 0 &  0.00 &  0.00 &  0.00 &  0.00 &  0.00 &  0.00 &  0.00 &  0.00 &  0.00 \\
 2 &  1.00 &  1.00 &  1.00 &  1.00 &  1.00 &  1.00 &  1.00 &  1.00 &  1.00 \\
 4 &  2.93 &  3.17 &  3.01 &  3.22 &  3.02 &  3.23 &  2.93 &  3.21 &  2.90 \\
 6 &  5.53 &  6.18 &  5.80 &  6.40 &  5.83 &  6.49 &  5.59 &  6.39 &  5.43 \\
 8 &  8.68 &  9.64 &  9.24 & 10.24 &  9.30 & 10.48 &  8.82 & 10.20 &  8.48 \\
10 & 12.28 & 13.24 & 13.21 & 14.43 & 13.30 & 14.92 & 12.52 & 14.35 & 12.03 \\
12 & 16.27 & 16.72 & 17.64 & 18.70 & 17.75 & 19.56 & 16.59 & 18.58 & 16.04 \\
14 & 20.60 & 19.95 & 22.47 & 22.87 & 22.57 & 24.18 & 20.96 & 22.68 & 20.51 \\
16 &       &       & 27.61 & 26.81 & 27.66 & 28.63 & 25.57 & 26.55 & 25.44 \\
18 &       &       &       &       & 33.21 & 32.83 & 30.33 & 30.11 & 30.80 \\
20 &       &       &       &       & 38.86 & 36.71 & 35.27 & 33.34 & 36.61 \\
   &       &       &       &       &       &       &       &       &       \\ 
$\beta_1$ & &      &       &       &       &       &       &       &       \\   
 0 &  5.2  &  6.1  &  5.6  &  5.5  &  5.5  &  5.3  &  4.9  &  5.1  &  5.6 \\
 2 &  6.5  &  6.9  &  6.7  &  6.3  &  6.6  &  6.1  &  6.0  &  5.9  &  7.5 \\
 4 &  8.7  &  8.5  &  8.4  &  8.1  &  8.5  &  8.0  &  7.9  &  7.7  & 10.7 \\
 6 & 11.8  & 10.9  & 10.8  & 10.7  & 11.1  & 10.7  & 10.4  & 10.3  & 14.8 \\
 8 &       &       & 13.7  & 13.9  & 14.3  & 14.0  & 13.5  & 13.5  & 19.4 \\
10 &       &       & 17.1  & 17.4  & 17.8  & 17.8  & 16.8  & 16.9  & 24.7 \\
12 &       &       & 20.7  & 21.0  & 21.3  & 21.7  &       &       & 30.5 \\
14 &       &       & 24.4  & 24.5  & 24.6  & 25.7  &       &       & 36.7 \\
16 &       &       &       &       & 28.4  & 29.6  &       &       & 43.5 \\
18 &       &       &       &       & 32.6  & 33.3  &       &       & 50.7 \\
20 &       &       &       &       & 37.8  & 36.7  &       &       & 58.4 \\
   &       &       &       &       &       &       &       &       &       \\
$\gamma_1$& &      &       &       &       &       &       &       &       \\    
 2 &  8.2  &  8.5  &  8.9  & 10.0  &  8.1  &  8.4  &  6.5  &  6.6  &  7.9 \\
 3 &  9.2  &  9.2  & 10.1  & 10.8  &  9.2  &  9.2  &  7.4  &  7.4  &  8.9 \\
 4 & 10.4  & 10.0  & 11.3  & 11.7  & 10.3  & 10.2  &  8.5  &  8.4  &  9.9 \\
 5 &       &       & 12.8  & 12.8  & 11.6  & 11.5  &  9.7  &  9.7  & 11.2 \\
 6 &       &       & 14.2  & 14.1  & 13.1  & 12.9  & 11.1  & 11.1  & 12.5 \\
 7 &       &       & 16.0  & 15.5  & 14.7  & 14.5  & 12.5  & 12.6  & 14.0 \\
 8 &       &       & 17.6  & 17.0  &       &       & 14.2  & 14.3  & 15.6 \\
 9 &       &       & 19.5  & 18.6  &       &       & 15.9  & 16.0  & 17.4 \\
10 &       &       &       &       &       &       & 17.8  & 17.8  & 19.2 \\
11 &       &       &       &       &       &       & 19.7  & 19.7  & 21.2 \\
12 &       &       &       &       &       &       & 21.8  & 21.5  & 23.3 \\
13 &       &       &       &       &       &       & 23.8  & 23.3  & 25.4 \\        
\hline
\end{tabular}
\end{table*}

\newpage 

\begin{table*}

\caption{Comparison of experimental data \cite{NDS} (upper line) for several $B(E2)$ ratios of $\gamma$-unstable nuclei
to predictions (lower line) by the Bohr Hamiltonian with $\beta$-dependent mass (with $\delta =\lambda =0$) for the Kratzer potential, for the
parameter values shown in Table I. See subsec. \ref{BE2gam} for further discussion. }

\bigskip

\begin{tabular}{l r@{.}l r@{.}l r@{.}l r@{.}l r@{.}l r@{.}l r@{.}l r@{.}l r@{.}l r@{.}l}

\hline
   \multicolumn{1}{l}{nucl.}
   &\multicolumn{2}{c} {$4_1\to 2_1 \over 2_1\to 0_1$}
    &\multicolumn{2}{c} {$6_1\to 4_1 \over 2_1\to 0_1$}
    &\multicolumn{2}{c} {$8_1\to 6_1 \over 2_1\to 0_1$}
   &\multicolumn{2}{c} {$10_1\to 8_1 \over 2_1\to 0_1$}
    &\multicolumn{2}{c} {$2_2 \to 2_1 \over 2_1\to 0_1$}
   &\multicolumn{2}{c}{$2_2 \to 0_1 \over 2_1\to 0_1$}
   &\multicolumn{2}{c}{$0_2 \to 2_1 \over 2_1\to 0_1$}
   &\multicolumn{2}{c}{$2_3 \to 0_1 \over 2_1 \to 0_1$}  \\
   
   & \omit\span & \omit\span & \omit\span & \omit\span &
  \omit\span &  \multicolumn{2}{c} {x $10^3$} &  \omit\span &
  \multicolumn{2}{c} {x $10^3$}
   \\

\hline 

$^{98}$Ru   & 1&44(25) & \omit\span & \omit\span & \omit\span &
1&62(61) &  36&0(152)      & \omit\span & \omit\span \\
           & 1&77 & 2&81 & 4&63 & 8&42 & 1&77 & 0&0 & 1&27 & 27&84 \\
           
$^{100}$Ru   & 1&45(13) & \omit\span & \omit\span & \omit\span &
0&64(12) &  41&1(52)      & 0&98(15) & \omit\span \\
           & 1&70 & 2&56 & 3&93 & 6&59 & 1&70 & 0&0 & 1&11 & 43&07 \\

$^{102}$Ru   & 1&50(24) & \omit\span & \omit\span & \omit\span &
0&62(7) &  24&8(7)      & 0&80(14) & \omit\span \\
           & 1&77 & 2&82 & 4&68 & 8&67 & 1&77 & 0&0 & 1&29 & 31&06 \\

$^{104}$Ru   & 1&18(28) & \omit\span & \omit\span & \omit\span &
0&63(15) &  35&0(84)      & 0&42(7) & \omit\span \\
           & 1&60 & 2&20 & 2&95 & 4&00 & 1&60 & 0&0 & 0&68 &25&59 \\

$^{102}$Pd   & 1&56(19) & \omit\span & \omit\span & \omit\span &
0&46(9) &  128&8(735)      & \omit\span & \omit\span \\
           & 1&63 & 2&31 & 3&25 & 4&77 & 1&63 & 0&0 & 0&87 & 41&64 \\
 
$^{104}$Pd   & 1&36(27) & \omit\span & \omit\span & \omit\span &
0&61(8) &  33&3(74)      & \omit\span & \omit\span \\
           & 1&70 & 2&52 & 3&74 & 5&83 & 1&70 & 0&0 & 0&99 & 24&16 \\  
           
 $^{106}$Pd   & 1&63(28) & \omit\span & \omit\span & \omit\span &
0&98(12) &  26&2(31)      & 0&67(18) & \omit\span \\
           & 1&74 & 2&66 & 4&13 & 6&83 & 1&74 & 0&0 & 1&12 & 22&91 \\
           
$^{108}$Pd   & 1&47(20) & 2&16(28) & 2&99(48) & \omit\span &
1&43(14) &  16&6(18)      & 1&05(13) & 1&90(29) \\
           & 1&66 & 2&38 & 3&42 & 5&11 & 1&66 & 0&0 & 0&89 & 30&31 \\ 
           
$^{110}$Pd   & 1&71(34) & \omit\span & \omit\span & \omit\span &
0&98(24) &  14&1(22)      & 0&64(10) & \omit\span \\
           & 1&60 & 2&18 & 2&94 & 4&06 & 1&60 & 0&0 & 0&75 & 42&18 \\ 
           
$^{106}$Cd   & 1&78(25) & \omit\span & \omit\span & \omit\span &
0&43(12) &  93&0(127)      & \omit\span & \omit\span \\
           & 1&66 & 2&37 & 3&34 & 4&76 & 1&66 & 0&0 & 0&83 & 16&97 \\
           
$^{108}$Cd   & 1&54(24) & \omit\span & \omit\span & \omit\span &
0&64(20) &  67&7(120)      & \omit\span & \omit\span \\
           & 1&67 & 2&40 & 3&43 & 5&01 & 1&67 & 0&0 & 0&87 & 19&88 \\
           
$^{110}$Cd   & 1&68(24) & \omit\span & \omit\span & \omit\span &
1&09(19) &  48&9(78)      & \omit\span & 9&85(595) \\
           & 1&85 & 3&14 & 5&63 &11&54 & 1&85 & 0&0 & 1&52 & 20&99 \\
           
$^{112}$Cd   & 2&02(22) & \omit\span & \omit\span & \omit\span &
0&50(10) &  19&9(35)      & 1&69(48) & 11&26(210) \\
           & 1&95 & 3&53 & 6&92 & 15&92 & 1&95 & 0&0 & 1&82 & 12&87 \\ 
           
$^{114}$Cd   & 1&99(25) & 3&83(72) & 2&73(97) & \omit\span &
0&71(24) &  15&4(29)      & 0&88(11) & 10&61(193) \\
           & 1&93 & 3&46 & 6&72 & 15&44 & 1&93 & 0&0 & 1&77 & 15&44 \\   
           
$^{116}$Cd   & 1&70(52) & \omit\span & \omit\span & \omit\span &
0&63(46) &  32&8(86)      & 0&02 & \omit\span \\
           & 1&69 & 2&47 & 3&52 & 5&05 & 1&69 & 0&0 & 0&90 & 10&02 \\  
           
$^{118}$Cd   & $>$1&85 & \omit\span & \omit\span & \omit\span &
\omit\span & \omit\span      & 0&16(4) & \omit\span \\
           & 1&70 & 2&51 & 3&65 & 5&41 & 1&70 & 0&0 & 0&95 & 13&14 \\  
           
$^{118}$Xe   & 1&11(7) & 0&88(27) & 0&49(20) & $>$0&73 &
\omit\span & \omit\span & \omit\span &\omit\span \\
           & 1&61 & 2&21 & 3&00 & 4&17 & 1&61 & 0&0 & 0&74 & 34&42 \\  
 
$^{120}$Xe   & 1&16(14) & 1&17(24) & 0&96(22) & 0&91(19) &
\omit\span & \omit\span & \omit\span &\omit\span \\
           & 1&56 & 2&06 & 2&64 & 3&43 & 1&56 & 0&0 & 0&62 & 42&25 \\            

$^{122}$Xe   & 1&47(38) & 0&89(26) & $>$0&44 & \omit\span &
   \omit\span & \omit\span    & \omit\span & \omit\span \\
           & 1&54 & 2&00 & 2&52 & 3&17 & 1&54 & 0&0 & 0&54 & 36&27 \\
           
$^{124}$Xe   & 1&34(24) & 1&59(71) & 0&63(29) & 0&29(8) &
0&70(19) &  15&9(46)      & \omit\span  & \omit\span \\
           & 1&55 & 2&03 & 2&57 & 3&25 & 1&55 & 0&0 & 0&53 & 28&40 \\
           
$^{128}$Xe   & 1&47(20) & 1&94(26) & 2&39(40) & 2&74(114) &
1&19(19) &  15&9(23)      & \omit\span  & \omit\span \\
           & 1&83   & 2&95 & 4&73 & 7&64 & 1&83 & 0&0 & 0&75 & 12&57 \\
  
$^{132}$Xe   & 1&24(18) & \omit\span & \omit\span & \omit\span &
1&77(29) &  3&4(7)      & \omit\span & \omit\span \\
           & 2&78 & 7&13 & 17&89 & 43&35 & 2&78 & 0&0 & 2&49 & 0&07 \\  
           
$^{130}$Ba   & 1&36(6) & 1&62(15) & 1&55(56) & 0&93(15) &
\omit\span  & \omit\span   & \omit\span  & \omit\span \\
           & 1&54 & 2&01 & 2&54 & 3&22 & 1&54 & 0&0 & 0&56 & 39&43 \\
           
$^{132}$Ba   & \omit\span & \omit\span & \omit\span & \omit\span &
3&35(64) &  90&7(177)      & \omit\span & \omit\span \\
           & 1&61 & 2&20 & 2&94 & 3&95 & 1&61 & 0&0 & 0&66 & 20&59 \\  
  
$^{134}$Ba   & 1&55(21) & \omit\span & \omit\span & \omit\span &
2&17(69) &  12&5(41)      & \omit\span & \omit\span \\
           & 2&13 & 4&10 & 7&88 & 15&19 & 2&13 & 0&0 & 1&26 & 6&22 \\    
  
$^{142}$Ba   & 1&40(17) & 0&56(14) & \omit\span & \omit\span &
\omit\span & \omit\span   & \omit\span & \omit\span \\
           & 1&54 & 1&99 & 2&46 & 3&04 & 1&54 & 0&0 & 0&45 & 21&34 \\

$^{148}$Nd   & 1&61(13) & 1&76(19) & \omit\span & \omit\span &
0&25(4) & 9&3(17)   & 0&54(6) & 32&82(816) \\
           & 1&57 & 2&08 & 2&67 & 3&47 & 1&57 & 0&0 & 0&59 & 30&88 \\ 
                        
\hline
\end{tabular}
\end{table*}  
  
\begin{table*}

\setcounter{table}{4} \caption{ (continued) }

\bigskip

\begin{tabular}{l r@{.}l r@{.}l r@{.}l r@{.}l r@{.}l r@{.}l r@{.}l r@{.}l r@{.}l r@{.}l}

\hline
   \multicolumn{1}{l}{nucl.}
   &\multicolumn{2}{c} {$4_1\to 2_1 \over 2_1\to 0_1$}
    &\multicolumn{2}{c} {$6_1\to 4_1 \over 2_1\to 0_1$}
    &\multicolumn{2}{c} {$8_1\to 6_1 \over 2_1\to 0_1$}
   &\multicolumn{2}{c} {$10_1\to 8_1 \over 2_1\to 0_1$}
    &\multicolumn{2}{c} {$2_2 \to 2_1 \over 2_1\to 0_1$}
   &\multicolumn{2}{c}{$2_2 \to 0_1 \over 2_1\to 0_1$}
   &\multicolumn{2}{c}{$0_2 \to 2_1 \over 2_1\to 0_1$}
   &\multicolumn{2}{c}{$2_3 \to 0_1 \over 2_1 \to 0_1$}  \\

   & \omit\span & \omit\span & \omit\span & \omit\span &
  \omit\span &  \multicolumn{2}{c} {x $10^3$} &  \omit\span &
  \multicolumn{2}{c} {x $10^3$}
   \\
           
\hline 
           
$^{152}$Gd   & 1&84(29) & 2&74(81) & \omit\span & \omit\span &
0&23(4) & 4&2(8)   & 2&47(78) & \omit\span \\
           & 1&80 & 2&96 & 5&14 & 10&30 & 1&80 & 0&0 & 1&41 & 32&70 \\
  
$^{154}$Dy   & 1&62(35) & 2&05(42) & 2&27(62) & 1&86(69) &
\omit\span   & \omit\span & \omit\span & \omit\span \\
           & 1&78 & 2&89 & 5&06 & 10&73 & 1&78 & 0&0 & 1&46 & 58&09 \\
  
$^{156}$Er   & 1&78(16) & 1&89(36) & 0&76(20) & 0&88(22) &
\omit\span   & \omit\span & \omit\span & \omit\span \\
           & 1&64 & 2&33 & 3&27 & 4&73 & 1&64 & 0&0 & 0&83 & 28&76 \\
  
$^{192}$Pt   & 1&56(12) & 1&23(55) & \omit\span & \omit\span &
1&91(16)   & 9&5(9) & \omit\span & \omit\span \\
           & 1&57 & 2&09 & 2&68 & 3&44 & 1&57 & 0&0 & 0&54 & 17&79 \\
            
 $^{194}$Pt   & 1&73(13) & 1&36(45) & 1&02(30) & 0&69(19) &
1&81(25) &  5&9(9)      & 0&01  & \omit\span \\
           & 1&56 & 2&07 & 2&63 & 3&34 & 1&56 & 0&0 & 0&52 & 19&45 \\
           
 $^{196}$Pt   & 1&48(3) & 1&80(23) & 1&92(23) & \omit\span &
\omit\span &  0&4      & 0&07(4)  & 0&06(6) \\
           & 1&61 & 2&21 & 2&97 & 4&04 & 1&61 & 0&0 & 0&69 & 23&11 \\ 
           
 $^{198}$Pt   & 1&19(13) & $>$1&78 & \omit\span & \omit\span &
 1&16(23) &  1&2(4)      & 0&81(22)  & 1&56(126) \\
           & 1&76 & 2&73 & 4&24 & 6&76 & 1&76 & 0&0 & 1&16 & 11&09 \\          
           
\hline
\end{tabular}
\end{table*}

\begin{table*}

\caption{Comparison of experimental data \cite{NDS} (upper line) for several $B(E2)$ ratios of axially symmetric prolate deformed nuclei
to predictions (lower line) by the Bohr Hamiltonian with $\beta$-dependent mass (with $\delta =\lambda =0$) for the Kratzer potential, for the
parameter values shown in Table II. In order to facilitate comparisons  
of the ${\rm N}=90$ isotones $^{150}$Nd, $^{152}$Sm, $^{154}$Gd, and $^{156}$Dy, to the predictions of the X(5) critical point symmetry
\cite{IacX5,BonX5,Bijker}, the relevant predictions are reported in the first line of the table, with the $\gamma_1\to gsb$ transitions
normalized to the $2_\gamma \to 0_1$ transition, which is set equal to 100, close to the average value for the first three $N=90$ isotones.  
See subsec. \ref{BE2def} for further discussion.}

\bigskip

\begin{tabular}{l r@{.}l r@{.}l r@{.}l r@{.}l r@{.}l r@{.}l r@{.}l r@{.}l r@{.}l r@{.}l}

\hline
   \multicolumn{1}{l}{nucl.}
   &\multicolumn{2}{c} {$4_1\to 2_1 \over 2_1\to 0_1$}
    &\multicolumn{2}{c} {$6_1\to 4_1 \over 2_1\to 0_1$}
    &\multicolumn{2}{c} {$8_1\to 6_1 \over 2_1\to 0_1$}
   &\multicolumn{2}{c} {$10_1\to 8_1 \over 2_1\to 0_1$}
    &\multicolumn{2}{c} {$2_\beta \to 0_1 \over 2_1\to 0_1$}
   &\multicolumn{2}{c}{$2_\beta \to 2_1 \over 2_1\to 0_1$}
   &\multicolumn{2}{c}{$2_\beta \to 4_1 \over 2_1\to 0_1$}
   &\multicolumn{2}{c}{$2_\gamma\to 0_1 \over 2_1 \to 0_1$}
   &\multicolumn{2}{c}{$2_\gamma\to 2_1 \over 2_1 \to 0_1$} 
   &\multicolumn{2}{c}{$2_\gamma\to 4_1 \over 2_1 \to 0_1$} \\

   & \omit\span & \omit\span & \omit\span & \omit\span &
  \multicolumn{2}{c} {x $10^3$} &  \multicolumn{2}{c} {x $10^3$} &  \multicolumn{2}{c} {x $10^3$} &
  \multicolumn{2}{c} {x $10^3$} &  \multicolumn{2}{c} {x $10^3$} &  \multicolumn{2}{c} {x $10^3$}
   \\

\hline 

X(5)         & 1&60 & 1&98 & 2&28 & 2&51 & 21&2 & 82&2 &  \multicolumn{2}{c}  {366} &
100&0 & 150&0 & 7&8 \\

$^{150}$Nd   & 1&52(4) & 1&84(14) & 2&05(13) & \omit\span &
4&4(8) & 61&7(98)      & \multicolumn{2}{c} {174(55)} &
26&1(22) & 49&6(26) & 14&8(98) \\
           & 1&55 & 1&98 & 2&55 & 3&40 & 36&4 & 93&5 &  \multicolumn{2}{c}  {443} &
95&9 & 150&0 & 9&0 \\

$^{152}$Sm   & 1&45(5) & 1&70(7) & 1&98(14) & 2&22(25) &
6&4(7) & 38&2(43)      & \multicolumn{2}{c} {132(15)} &
25&1(17) & 64&6(48) & 5&4(5) \\
           & 1&55 & 1&98 & 2&58 & 3&53 & 49&1 & 113&4 &  \multicolumn{2}{c}  {475} &
84&0 & 130&9 & 7&8 \\

$^{154}$Sm   & 1&40(5) & 1&67(7) & 1&83(11) & 1&81(11) &
5&4(13) &  \omit\span      & \multicolumn{2}{c} {25(6)} &
18&4(34) &  \omit\span & 3&9(7) \\
           & 1&46 & 1&67 & 1&86 & 2&05 & 24&7 & 45&7 &  \multicolumn{2}{c}  {136} &
47&8 & 69&9 & 3&7 \\

$^{154}$Gd   & 1&56(7) & 1&82(11) & 1&99(12) & 2&29(27) &
5&5(5) & 42&7(41)      & \multicolumn{2}{c} {125(11)} &
36&3(34) & 78&3(69) & 11&0(10) \\
           & 1&55 & 1&98 & 2&57 & 3&54 & 55&4 & 122&5 &  \multicolumn{2}{c}  {486} &
114&7 & 175&6 & 10&1 \\

$^{156}$Gd & 1&41(5) & 1&58(6) & 1&71(10) & 1&68(9) &  3&4(3) &
\multicolumn{2}{c} {18(2)} & \multicolumn{2}{c} {22(2)} &
25&0(15) & 38&7(24) & 4&1(3)  \\
           & 1&47 & 1&69 & 1&90 & 2&13 & 30&8 & 56&5 & \multicolumn{2}{c} {166} &
70&7 & 103&3  & 5&4  \\

$^{158}$Gd & 1&46(5) &  \omit\span        & 1&67(16) & 1&72(16) &
1&6(2) & 0&4(1) & 7&0(8) &
17&2(20) & 30&3(45) & 1&4(2) \\
           & 1&45 & 1&66 & 1&82 & 1&98 & 24&1 & 42&8 & \multicolumn{2}{c} {119} &
63&9 & 92&6 &  4&8 \\

$^{156}$Dy   & 1&75(14) & 1&34(12) & 1&94(13) & 2&45(21) &
\omit\span & \omit\span      & \omit\span &
48&2(35) & 63&0(78) & 84&4(141) \\
           & 1&56 & 2&03 & 2&70 & 3&83 & 53&2 & 124&3 &  \multicolumn{2}{c}  {531} &
151&8 & 232&6 & 13&4 \\

$^{158}$Dy & 1&45(10) & 1&86(12) & 1&86(38) & 1&75(28) &
\multicolumn{2}{c} {12(3)} & \multicolumn{2}{l} {19(4)} &
\multicolumn{2}{c} {66(16)} &
32&2(78) & 103&8(258) & 11&5(48) \\
           & 1&48 & 1&73 & 1&98 & 2&28 & 32&5 & 63&0 & \multicolumn{2}{c} {202} &
97&9 & 143&6  &  7&6 \\

$^{160}$Dy & 1&46(7) & 1&23(7) & 1&70(16) & 1&69(9) &  3&4(4) &
\omit\span   & 8&5(10) &
23&2(21) & 43&8(42) & 3&1(3) \\
           & 1&46 & 1&66 & 1&83 & 2&00 & 23&5 & 42&5 & \multicolumn{2}{c} {122} &
87&4 & 126&6 & 6&5 \\

$^{162}$Dy & 1&45(7) & 1&51(10) & 1&74(10) & 1&76(13) & \omit\span
&  \omit\span   &  \omit\span   &
0&12(1) & 0&20  & 0&02\\
           & 1&45 & 1&65 & 1&80 & 1&95 & 23&7 & 41&4 & \multicolumn{2}{c} {112} &
92&3 & 133&2 & 6&8 \\

$^{164}$Dy & 1&30(7) & 1&56(7) & 1&48(9) & 1&69(9) &  \omit\span &
\omit\span &  \omit\span   &
19&1(22) & 38&3(39) & 4&6(5) \\
           & 1&45 & 1&64 & 1&79 & 1&93 & 23&6 & 40&6 & \multicolumn{2}{c} {107} &
100&4 & 144&7  & 7&4  \\

$^{162}$Er &  \omit\span         &  \omit\span                &
\omit\span  & \omit\span     & \multicolumn{2}{c} {8(7)} &
\omit\span  & \multicolumn{2}{c} {170(90)} &
32&5(28) & 77&0(56) &  9&4(69) \\
           & 1&48 & 1&73 & 1&97 & 2&25 & 28&9 & 57&1 & \multicolumn{2}{c} {189} &
100&4 & 147&1 & 7&8 \\

$^{164}$Er & 1&18(13) &   \omit\span        & 1&57(9) & 1&64(11) &
\omit\span &  \omit\span  &  \omit\span   &
23&9(35) & 52&3(72) & 7&8(12) \\
           & 1&46 & 1&67 & 1&86 & 2&05 & 27&0 & 49&0 & \multicolumn{2}{c} {141} &
110&5 & 160&2 & 8&2 \\

$^{166}$Er & 1&45(12) & 1&62(22) & 1&71(25) & 1&73(23) &
\omit\span    &  \omit\span   &  \omit\span   &
25&7(31) & 45&3(54) & 3&1(4) \\
           & 1&48 & 1&74 & 2&00 & 2&31 & 21&2 & 39&2 & \multicolumn{2}{c} {117} &
\omit\span  &   \omit\span  &  \omit\span   \\  

$^{168}$Er & 1&54(7) & 2&13(16) & 1&69(11) & 1&46(11) & \omit\span
&  \omit\span   &  \omit\span   &
23&2(15) & 41&1(31) & 3&0(3) \\
           & 1&45 & 1&64 & 1&78 & 1&92 & 27&6 & 46&2 & \multicolumn{2}{c} {116} &
100&6 & 144&9 &  7&4  \\

$^{170}$Er &  \omit\span         &  \omit\span         & 1&78(15)
& 1&54(11) & 1&4(1) & 0&2(2) & 6&8(12) &
17&7(9) &  \omit\span        & 1&4(4) \\
           & 1&46 & 1&66 & 1&83 & 2&01 & 42&8 & 70&7 & \multicolumn{2}{c} {173} &
84&6 & 122&2 & 6&3  \\

$^{166}$Yb & 1&43(9) & 1&53(10) & 1&70(18) & 1&61(80) & \omit\span
&  \omit\span   &  \omit\span   &
     \omit\span    &  \omit\span    &  \omit\span        \\
           & 1&48 & 1&73 & 1&97 & 2&27 &  35&4 & 66&7 & \multicolumn{2}{c} {206} &
115&2 & 168&2 & 8&8 \\

$^{168}$Yb &  \omit\span         &   \omit\span             &
\omit\span  & \omit\span       & 8&6(9) &  \omit\span   &
\omit\span   &
22&0(55) & 45&9(73) & 8&6 \\
           & 1&46 & 1&68 & 1&86 & 2&06 & 26&3 & 48&2 & \multicolumn{2}{c} {142} &
87&6 & 127&2 & 6&6  \\


$^{170}$Yb &  \omit\span         &  \omit\span        & 1&79(16) &
1&77(14) & 5&4(10) &  \omit\span   &  \omit\span   &
13&4(34) & 23&9(57) & 2&4(6) \\
           & 1&47 & 1&69 & 1&90 & 2&13 & 31&9 & 58&0 & \multicolumn{2}{c} {168} &
69&4 & 101&3 & 5&3  \\

$^{172}$Yb & 1&42(10) & 1&51(14) & 1&89(19) & 1&77(11) & 1&1(1) &
3&7(6) & \multicolumn{2}{c} {12(1)} & 6&3(6) & \omit\span   & 0&6(1) \\
           & 1&46 & 1&66 & 1&83 & 2&01 & 29&6 & 51&6 & \multicolumn{2}{c} {139} &
51&0 & 74&2 & 3&8  \\

$^{174}$Yb & 1&39(7) & 1&84(26) & 1&93(12) & 1&67(12) & \omit\span
&  \omit\span  &  \omit\span  &
    \omit\span  & 12&4(29)   & \omit\span          \\
           & 1&44 & 1&62 & 1&74 & 1&86 & 20&6 & 34&3 & \multicolumn{2}{c} {85} &
45&7 & 65&8 & 3&4  \\

$^{176}$Yb & 1&49(15) & 1&63(14) & 1&65(28) & 1&76(18) &
\omit\span    &  \omit\span   &  \omit\span   &
9&8
 &  \omit\span   &  \omit\span       \\
           & 1&45 & 1&65 & 1&81 & 1&97 & 28&6 & 49&0 & \multicolumn{2}{c} {128} &
64&5 & 93&4 & 4&8  \\

\hline
\end{tabular}
\end{table*}

\begin{table*}
\setcounter{table}{5} \caption{ (continued) }

\bigskip

\begin{tabular}{l r@{.}l r@{.}l r@{.}l r@{.}l r@{.}l r@{.}l r@{.}l r@{.}l r@{.}l r@{.}l}

\hline
   \multicolumn{1}{l}{nucl.}
   &\multicolumn{2}{c} {$4_1\to 2_1 \over 2_1\to 0_1$}
    &\multicolumn{2}{c} {$6_1\to 4_1 \over 2_1\to 0_1$}
    &\multicolumn{2}{c} {$8_1\to 6_1 \over 2_1\to 0_1$}
   &\multicolumn{2}{c} {$10_1\to 8_1 \over 2_1\to 0_1$}
    &\multicolumn{2}{c} {$2_\beta \to 0_1 \over 2_1\to 0_1$}
   &\multicolumn{2}{c}{$2_\beta \to 2_1 \over 2_1\to 0_1$}
   &\multicolumn{2}{c}{$2_\beta \to 4_1 \over 2_1\to 0_1$}
   &\multicolumn{2}{c}{$2_\gamma\to 0_1 \over 2_1 \to 0_1$}
   &\multicolumn{2}{c}{$2_\gamma\to 2_1 \over 2_1 \to 0_1$} 
   &\multicolumn{2}{c}{$2_\gamma\to 4_1 \over 2_1 \to 0_1$} \\

   & \omit\span & \omit\span & \omit\span & \omit\span &
  \multicolumn{2}{c} {x $10^3$} &  \multicolumn{2}{c} {x $10^3$} &  \multicolumn{2}{c} {x $10^3$}
   & \multicolumn{2}{c} {x $10^3$} &  \multicolumn{2}{c} {x $10^3$} &  \multicolumn{2}{c} {x $10^3$}
   \\
   
\hline 

$^{174}$Hf &  \omit\span            &   \omit\span       &
\omit\span &  \omit\span  & \multicolumn{2}{c}{14(4)} & \omit\span
& \multicolumn{2}{c} {9(3)} &
31&6(161) & 48&7(124) & \omit\span        \\
           & 1&49 & 1&76 & 2&05 & 2&42 & 47&1 & 87&5 & \multicolumn{2}{c}{264} &
69&7 & 102&9 & 5&5  \\

$^{176}$Hf & \omit\span  & \omit\span      & \omit\span       &
\omit\span & 5&4(11) & \omit\span & \multicolumn{2}{c} {31(6)} &
21&3(26) & \omit\span  &   \omit\span     \\
           & 1&46 & 1&68 & 1&86 & 2&06 & 29&1 & 52&2 & \multicolumn{2}{c} {148} &
60&3 & 87&8 & 4&6 \\

$^{178}$Hf & \omit\span       & 1&38(9) & 1&49(6) & 1&62(7) &
0&4(2) & \omit\span & 2&4(9) &
24&5(39) & 27&7(28) & 1&6(2) \\
           & 1&46 & 1&68 & 1&86 & 2&06 & 27&1 & 49&2 & \multicolumn{2}{c} {142} &
75&7 & 110&0 & 5&7 \\

$^{180}$Hf & 1&48(20) & 1&41(15) & 1&61(26) & 1&55(10) &
\omit\span & \omit\span & \omit\span &
24&5(47) & 32&9(56)  & \omit\span        \\
           & 1&46 & 1&66 & 1&83 & 2&00 & 34&6 & 58&6 & \multicolumn{2}{c} {150} &
80&3 & 116&0 & 6&0 \\

$^{182}$W  & 1&43(8) & 1&46(16) & 1&53(14) & 1&48(14) &  6&6(6) &
4&6(6) & \multicolumn{2}{c} {13(1)} &
24&8(12) & 49&2(24) & 0&2\\
           & 1&46 & 1&67 & 1&85 & 2&05 & 33&0 & 57&5 & \multicolumn{2}{c} {155} &
82&0 & 118&9 & 6&1  \\

$^{184}$W  & 1&35(12) & 1&54(9) & 2&00(18) & 2&45(51) & 1&8(3) &
\omit\span & \multicolumn{2}{c} {24(3)} &
37&1(28) & 70&6(51) & 4&0(4) \\
           & 1&48 & 1&73 & 1&97 & 2&27 &  38&9 & 71&8 & \multicolumn{2}{c} {214} &
128&4 & 187&1 & 9&8 \\

$^{186}$W  & 1&30(9) & 1&69(12) & 1&60(12) & 1&36(36) & \omit\span & \omit\span
& \omit\span & 41&7(92) & 91&0(201) & \omit\span        \\
           & 1&49 & 1&77 & 2&08 & 2&48 & 47&3 & 89&1 & \multicolumn{2}{c} {275} &
174&0 & 254&4 & 13&3 \\

$^{186}$Os & 1&45(7) & 1&99(7) & 1&89(11) & 2&06(44) & \omit\span
& \omit\span & \omit\span &
109&4(71) & 254&6(150) & 13&0(47) \\
           & 1&50 & 1&81 & 2&16 & 2&63 & 39&2 & 81&0 & \multicolumn{2}{c} {288} &
173&5 & 255&9 & 13&6 \\

$^{188}$Os & 1&68(11) & 1&75(11) & 2&04(15) & 2&38(32)        &
\omit\span & \omit\span & \omit\span &
63&3(92) & 202&5(304) & 43&0(74) \\
           & 1&52 & 1&87 & 2&29 & 2&87 & 33&6 & 78&5 & \multicolumn{2}{c} {330} &
246&6 & 366&2 & 19&7 \\

$^{230}$Th & 1&36(8)         & \omit\span      & \omit\span &
\omit\span & 5&7(26) & \omit\span & \multicolumn{2}{c} {20(11)} &
15&6(59) & 28&1(100) & 1&8(11) \\
           & 1&46 & 1&68 & 1&86 & 2&07  & 31&4 & 55&6 & \multicolumn{2}{c} {155} &
66&9 & 97&3 & 5&1 \\

$^{232}$Th & 1&44(15) & 1&65(14) & 1&73(12) & 1&82(15) &
\multicolumn{2}{c}{14(6)} & 2&6(13) & \multicolumn{2}{c} {17(8)} &
14&6(28) & 36&4(56) & 0&7 \\
           & 1&45 & 1&65 & 1&80 & 1&96 & 26&0 & 44&9 & \multicolumn{2}{c}{119} &
61&9 & 89&6 & 4&6 \\

$^{234}$U  & \omit\span &   \omit\span     & \omit\span
&\omit\span &\omit\span &\omit\span & \omit\span &
12&5(27) & 21&1(44) & 1&2(3) \\
           & 1&45 & 1&63 & 1&76 & 1&88 & 19&9 & 33&8 & \multicolumn{2}{c}{87} &
44&7 & 64&5 & 3&3  \\

$^{236}$U  & 1&42(11) & 1&55(11) & 1&59(17) & 1&46(17) &
\omit\span & \omit\span & \omit\span & \omit\span
      &   \omit\span    &   \omit\span    \\
           & 1&44 & 1&62 & 1&75 & 1&86 & 19&5 & 32&7 & \multicolumn{2}{c}{82} &
45&5 & 65&6 & 3&3 \\

$^{238}$U  & \omit\span       & \omit\span       & 1&45(23) &
1&71(22) & 1&4(6) & 3&6(14) & \multicolumn{2}{c}{12(5)} &
10&8(8) & 18&9(17)  & 1&2(1) \\
           & 1&44 & 1&62 & 1&74 & 1&85 & 19&3 & 32&3 & \multicolumn{2}{c}{80} &
39&4 & 56&8 & 2&9 \\

$^{238}$Pu &  \omit\span            &  \omit\span     & \omit\span
& \omit\span& \multicolumn{2}{c}{14(4)} &\omit\span &
\multicolumn{2}{c}{11(4)} &
   \omit\span   &  \omit\span   &  \omit\span      \\
           & 1&44 & 1&61 & 1&73 & 1&82 & 18&8 & 30&8 &  \multicolumn{2}{c}{74} &
42&2 & 60&7 & 3&1 \\

$^{250}$Cf &  \omit\span            &  \omit\span     & \omit\span
&\omit\span & \omit\span&\omit\span &\omit\span &
6&8(17) & 10&9(25) & 0&6(1) \\
           & 1&45 & 1&65 & 1&80 & 1&96 & 15&5 & 26&1 & \multicolumn{2}{c}{66} &
  \omit\span  &   \omit\span  &  \omit\span   \\  
  
\hline
\end{tabular}
\end{table*}

\begin{table}

\caption{Numerical values of the rescaling parameter $A$ for the $\gamma$-unstable nuclei shown in Figs. 2 and 3.
$\beta_0$ is the position of the minimum of the effective potential, calculated using in Eq. (\ref{b0unst}) the parameter values given in Table I, 
$\beta_{exp}$ is the value of the quadrupole deformation taken from the experimental value of $B(E2;0_1\to 2_1)$ \cite{Raman}, 
while $A$ is calculated from Eq. (\ref{Aunst}). }

\bigskip

\begin{tabular}{ r r l r}
\hline 
 & & & \\
nucleus & $\beta_0$ & $\beta_{exp}$ & $10^{3}A$ \\

\hline

$^{126}$Xe & 123.73 & 0.1881 & 1.520 \\
$^{128}$Xe &  66.00 & 0.1836 & 2.782 \\
$^{130}$Xe &  60.17 & 0.169  & 2.809 \\
$^{132}$Xe &  20.00 & 0.1409 & 7.045 \\
$^{134}$Xe &  12.00 & 0.119  & 9.917 \\

$^{130}$Ba & 286.86 & 0.2183 & 0.761 \\
$^{132}$Ba & 103.53 & 0.186  & 1.797 \\
$^{134}$Ba &  38.00 & 0.1609 & 4.234 \\
$^{136}$Ba &  18.01 & 0.1258 & 6.983 \\

\hline
\end{tabular}
\end{table}

\begin{table}

\caption{Numerical values of the rescaling parameter $A$ for the prolate deformed nuclei shown in Figs. 4, 5 and 6.
$\beta_0$ is the position of the minimum of the effective potential, calculated using in Eq. (\ref{b0prol}) the parameter values given in Table II, 
$\beta_{exp}$ is the value of the quadrupole deformation taken from the experimental value of $B(E2; 0_1\to 2_1)$ \cite{Raman}, 
while $A$ is calculated from Eq. (\ref{Aprol}). }

\bigskip

\begin{tabular}{ r r l r}
\hline 
 & & & \\
nucleus & $\beta_0$ & $\beta_{exp}$ & $10^{3}A$ \\

\hline

$^{150}$Nd &  90.94 & 0.2853 & 3.137 \\
   
$^{152}$Sm & 139.49 & 0.3064 & 2.197 \\

$^{154}$Gd & 164.23 & 0.3120 & 1.900 \\
$^{156}$Gd & 365.59 & 0.3378 & 0.924 \\
$^{158}$Gd & 449.05 & 0.3484 & 0.776 \\
$^{160}$Gd & 611.70 & 0.3534 & 0.578 \\

$^{156}$Dy & 131.89 & 0.2929 & 2.221 \\
$^{158}$Dy & 260.05 & 0.3255 & 1.252 \\
$^{160}$Dy & 383.71 & 0.3387 & 0.883 \\
$^{162}$Dy & 527.88 & 0.3430 & 0.650 \\
$^{164}$Dy & 584.40 & 0.3481 & 0.596 \\

\hline
\end{tabular}
\end{table}

\end{document}